\documentclass[a4paper,12pt]{article}
\pdfoutput=1
\usepackage{jheppub}
\usepackage{graphicx}
\usepackage{amssymb}
\usepackage{amsmath}
\usepackage{amsfonts}
\usepackage{hyperref}
\usepackage{xcolor}

	\def\bea{\begin{eqnarray}}
	\def\eea{\end{eqnarray}}
	\def\be{\begin{equation}}
	\def\ee{\end{equation}}
	\def\nn{\nonumber\\}

	\def\mm{\mathcal}
	
	\def\pa{\partial}

	\definecolor{db}{rgb}{0,0.08,0.45}
	\definecolor{brick}{rgb}{0.6,0.1,0.3}
	
	\definecolor{zz}{rgb}{1,0,0}
	\definecolor{zz2}{rgb}{0.7,0.1,0.1}
	\definecolor{yy}{rgb}{0.05,0.9,0.05}
	\definecolor{ww}{rgb}{0.6,0.1,0.3}
	
	\definecolor{rr}{cmyk}{0,0,0,1}
	\definecolor{vv}{rgb}{0.5,0,0.5}
	\definecolor{ss}{cmyk}{0,0,0,1}
	
	\def\zz{\textcolor{zz}}

	\definecolor{brick}{rgb}{0.5,0,0.5}

	\def\a{\alpha} \def\b{\beta}  \def\d{\delta}   \def\z{\zeta} \def\n{\nu} \def\m{\mu} \def\s{\sigma} \def\r{\rho}   \def\l{\lambda}  
	\def\G{\Gamma}  \def\D{\Delta}   \def\L{\Lambda}   \def\O{\Omega}

\title{Loops in AdS: From the Spectral Representation to Position Space III}

\author[\Psi]{Dean Carmi}
\affiliation[\Psi]{Department of Mathematics and Physics University of Haifa at Oranim, Kiryat Tivon 36006, Israel}
\emailAdd{deancarmi1@gmail.com}

\abstract{
We study loop amplitudes in anti de-Sitter space via the spectral representation. We consider loops of spinning fields and in particular gauge fields, and derive various identities connecting different families of loop diagrams, at different number of loops, different spins, different masses. Such identities are useful for the computation of Witten diagrams. Considering the theory of large-$N_f$ conformal scalar QED defined on AdS space, we derive an analytic expression for the exact 4-point correlation function at sub-leading order in $\frac{1}{N_f}$. Additionally, we derive analytic expressions for bulk 2-point functions and boundary 4-point functions for various families of diagrams, which we denote as ``blob diagrams''. Finally we study 4-point ladder diagrams with spinning fields, and we derive integral expressions for the spectral representation of a $k$-loop ladder diagram.}

\begin{document} 
\maketitle
\flushbottom

\section{Introduction}
\label{sec:se0}

The importance of anti de-Sitter space-time in physics arises primarily due its role in the AdS/CFT correspondence. The latter is a duality that relates a gravitational theory in AdS with a (non-gravitational) quantum field theory (QFT) theory on the boundary of AdS. Deep insights into quantum gravity have been obtained by studying observables in the boundary conformal field theory (CFT). Vice versa, insights into strongly coupled QFT have been obtained by perturbatively studying the bulk AdS theory.

In fact, the bulk theory in AdS space does not have to be a gravitational theory, and one can consider a non-gravitational QFT defined on AdS. If there is no gravity in the bulk, the boundary theory is then a non-local conformal theory  (i.e there is no stress tensor), and the boundary correlators are constrained by conformal symmetry. We define our theory in the bulk by specifying the matter content and Lagrangian for the bulk fields in AdS. Important physical observables in the theory are AdS scattering amplitudes, which can be obtained perturbatively via the computation of Feynman diagrams in AdS. Feynman diagrams in AdS are computed in a similar fashion to Feynman diagrams in Minkowski space, but the propagators are AdS propagators. Because there is no full translational invariance in AdS space, momentum-space techniques are less powerful compared to flat space. 

 The Feynman diagrams with all legs ending on the boundary of AdS define conformal correlators for the boundary conformal theory. AdS Feynman diagrams (often known as Witten diagrams) are generally harder to compute compared to their flat-space companions. Indeed, analytical results for loop diagrams in AdS are scarce. Various techniques have been employed to perform computations of Witten diagrams: in position space \cite{Liu:1998th,Liu:1998ty,Dolan:2000ut,Freedman:1998tz,DHoker:1998ecp,Freedman:1998bj,DHoker:1998bqu,DHoker:1999mqo,Zhou:2018sfz}, momentum-space \cite{Raju:2010by,Raju:2011mp,Albayrak:2019asr,Albayrak:2018tam,Albayrak:2020fyp,Albayrak:2020isk}, Mellin-space \cite{Penedones:2010ue,Fitzpatrick:2011ia,Paulos:2011ie,Rastelli:2017udc,Rastelli:2016nze,Cardona:2017tsw,Yuan:2017vgp,Yuan:2018qva}, spectral techniques \cite{Carmi:2018qzm,Carmi:2019ocp,Carmi:2021dsn,Ankur:2023lum}, using conformal or super symmetry \cite{Aharony:2016dwx,Henriksson:2017eej,Alday:2017xua,Alday:2017gde,Mazac:2018mdx,Carmi:2020ekr,Caron-Huot:2018kta,Alday:2017vkk,Aprile:2017bgs,Aprile:2017qoy,Bissi:2020woe,Bissi:2020wtv}. For other works on loop amplitudes in AdS see \cite{,Fitzpatrick:2010zm,Fitzpatrick:2011hu,Ponomarev:2019ofr,Bertan:2018khc,Bertan:2018afl,Beccaria:2019stp,Fitzpatrick:2011dm,Giombi:2017hpr,Costantino:2020vdu,Antunes:2020pof,Meltzer:2019nbs,Nagaraj:2020sji,Meltzer:2020qbr,Albayrak:2020bso,Zhou:2020ptb,Eberhardt:2020ewh}.

\subsection{Summary of results}

In the current work we use the spectral representation in AdS to study and compute AdS amplitudes at the loop level. We consider both scalar and spinning fields in AdS, and study a family of diagrams which we call ``blob diagrams", see Fig~\ref{fig:bubblerelations161}. Blob diagrams are particularly susceptible to spectral techniques, which is the main reason that we study them, see also \cite{Carmi:2019ocp,Carmi:2021dsn}.
We derive a set of vertex/propagator identities that relate various diagrams, and these identities enable to compute certain "blob diagrams" in terms of lower loop (and often tree-level) diagrams. Other identities enable to relate diagrams across different masses, dimensions, and spins. In previous work \cite{Ankur:2023lum} we initiated a study of large-$N_F$ scalar QED on $AdS$. The spectral representation enables to resum bubble diagrams in the bulk, which gives the exact photon propagator at order $1/N_f$. This then enables to compute the 4-point functions. We focus here on the bulk conformal point of scalar QED, in which we obtain an analytical expression for the (non-perturbative) 4-point correlator, in terms of a tree-level exchange diagram.

In section~\ref{sec:se1} we review the spectral representation of bulk 2-point functions of spinning fields in AdS, and of boundary 4-point functions. We discuss the 1-loop bubble diagram in the spectral representation, and define the family of ``Blob diagrams". In section~\ref{sec:se2} we derive the following vertex/propagator identities:

\begin{itemize}

\item \underline{Identity 0:} Starting from a 4-point blob diagram with\footnote{Here we defined $a\equiv \frac{\D_1-\D_2}{2}$, $b\equiv \frac{\D_3-\D_4}{2}$ to be the differences of scaling dimensions of the external operators of the blob diagram.} with a given $a$ and $b$, this identity enables to raise and lower $a$ and $b$ by any integer numbers. The identity thus relates an infinite tower of different blob diagrams together. The identity holds for space time dimensions $d=2,4$, and any spin $l$ for the exchanged operator.

\item \underline{Identity 1:} Starting from a 4-point blob diagram with a given external $\D_i$, $i=1, \ldots, 4$. This identity states that the expression for the blob diagram equals that of a blob diagram with external $\D_i'$, where $\D_i'$ are defined in Eq.~\ref{eq:sdbsmd7s}.  The identity holds for any $d$ and $l$.

\item \underline{Identity 2:} Starting from a 4-point blob diagram with\footnote{Here we defined $a_p\equiv \frac{\D_1+\D_2}{2}$, $b_p\equiv \frac{\D_3+\D_4}{2}$.} with a given $a_p$ and $b_p$, this identity enables to raise and lower $a_p$ and $b_p$ by any integer numbers. The identity thus relates an infinite tower of different blob diagrams together. The identity holds for any $d$ and any spin $l$.

\item \underline{Identity 3:} is a vertex/propagator identity, relating diagrams with different vertices. It holds for any $d$ and spin $l$.

\item \underline{Identity 4:} Relates loop sunset diagrams with shifted scaling dimensions, and bubble diagrams. This identity holds for scalars and fermions in $AdS_3$.

\item \underline{Identity 6:} is a vertex/bubble identity, relating a vertex with and without a 1-loop bubble. This identity enables to reduce certain loop diagrams to lower loop diagrams, sometimes to tree level diagrams. The identity is specific for $l=1$ and $\D=1$.

\end{itemize}

 In section~\ref{sec:se3} we study the 4-point correlator in the large-$N_f$ scalar QED on $AdS_3$. At leading order in $\frac{1}{N_f}$, the exact propagator is given by a resummation of bubble diagrams, and the 4-point function is given by an exchange diagram of the exact photon propagator. Remarkably, we show that this exact 4-point function is in fact equal to a certain tree level exchange diagram. In section~\ref{sec:se4} we compute bulk AdS 2-point correlators of spinning fields, in particular focusing on spin-1 bulk correlators. We find a relation between 2-point bulk correlators and 4-point boundary correlators. In section~\ref{sec:se6} we compute various 4-point 1-loop bubble diagrams for scalars and fermions. We connect the (double) discontinuities of the bubble diagram to tree level contact diagrams. In section~\ref{sec:se7} we go beyond ``blob diagrams" and consider 4-point ladder diagrams of spinning fields in the spectral representation. We derive expressions for these diagrams in terms of integrals of products of 6J-symbols.  In section~\ref{sec:se8} we summarise the results and discuss future directions. In appendix~\ref{sec:A1} we show details of the computation of ladder diagrams done in section~\ref{sec:se7} . In appendix~\ref{sec:B1} we write blob diagrams in terms of Legendre functions, which suggests a way to compute them in position space. In appendix~\ref{sec:se5} we show additional relations for 4-point blob diagrams. In section~\ref{sec:c1} we prove a relation between blob diagrams with different spins $l$, and in section~\ref{sec:c2} we prove a relation between blob diagrams with different space-time dimension $d$.

\subsection{QFT on AdS}

\begin{figure}[t]
\centering
\includegraphics[clip,height=5.5cm]{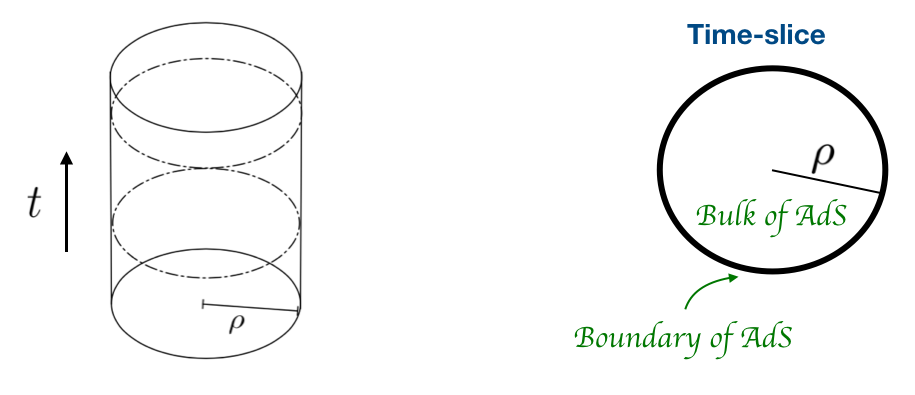}
\caption{Anti de-Sitter space-time. \textbf{Left:} The AdS cylinder in global coordinates. \textbf{Right:}  A time-slice of the AdS cylinder. The circle is the boundary of AdS, and the inside of the circle is the bulk of AdS.}
\label{fig:1324}
\end{figure}

Anti de-Sitter space-time is a maximally symmetric space-time, which is a solution of Einstein's equations with negative cosmological constant. $AdS_{d+1}$ is a hyperboloid inside flat space time:
\begin{align} 
X_0^2+X_{d+1}^2- \sum^{d}_{i=1} X_i^2 = R_{AdS}^2
\end{align} 
where $R_{AdS}$ is the radius of AdS. The metric of AdS space-time in global coordinates is:
\begin{align} 
ds^2 = \frac{1}{\cos^2 \big( \frac{\r}{R_{AdS}} \big)} \Big[ dt^2 -d\r^2-\sin^2  \big( \frac{\r}{R_{AdS}} \big) d\O_{d-1}^2  \Big]
\end{align} 
Where $\r$ is an AdS radial coordinate, and the angular coordinates $\O_{d-1}$ parametrise a $(d-1)$-sphere. The AdS cylinder is shown in the LHS of Fig.~\ref{fig:1324}, and a time-slice of this cylinder is shown in the RHS of Fig.~\ref{fig:1324}.

Now consider any quantum field theory defined via a Lagrangian $\mm{L}$. Instead of considering this QFT on flat-space $\mathbb R^{d+1}$, we consider the theory on a fixed curved $AdS_{d+1}$ background. The action is given by $S=\int_{AdS} \sqrt{g} \mm{L}$, and the AdS partition function is then $Z= \int \mm{D} \phi e^{-S[\phi]}$. For example, the Euclidean action of a scalar field of mass $m$ and cubic coupling $\l$ is given by:
\begin{align} 
\label{eq:bksd8d}
S=\int_{AdS_{d+1}} d^{d+1}x \sqrt{g} \Big( \frac{1}{2} (\pa \phi)^2 +m^2\phi^2 +\frac{\l}{3!} \phi^3 \Big)
\end{align} 
This implies the equation of motion:
\begin{align} 
g^{\m \n} \nabla_\m  \nabla_\n \phi -m^2\phi -\frac{\l}{3!}\phi^3=0
\end{align} 
The bulk-to-bulk propagator between two points $x_1$ and $x_2$ in the bulk of AdS is the Green's function, obeying the equation:
\begin{align} 
\big(\pa^2_{x_1}+m^2 \big)G_\D(x_1,x_2) =(2\pi)^{d+1} \d^{d+1}_{AdS}(x_1-x_2)
\end{align} 
where $\D$ is related to the mass as $R_{AdS}^2m^2= \D(\D-d)$. The bulk-to-bulk propagator has the explicit expression:
\begin{align} 
\label{eq:fkjdf3}
G_\D(x_1,x_2) =  \frac{\G_\D}{2\pi^{\frac{d}{2}} \G_{\D-\frac{d}{2}+1}} \zeta^{-\D} {}_2F_1 (\D,\D-\frac{d-1}{2},2\D-d+1,-4\zeta^{-1})
\end{align} 
Here  $\zeta= \frac{(z_1-z_2)^2+(\vec{x}_1-\vec{x}_2)^2}{z_1z_2}$ is the chordal distance squared in AdS, $\vec{x}_i$ are boundary directions, and $z_i$ are AdS radial coordinates. The bulk-to-boundary propagator is obtained by taking one of the points of the bulk-to-bulk propagator to the AdS boundary:
\begin{align} 
K_{\D}(P_1,x_2)= G_\D(x_1,x_2) \Big|_{x_1 \to P_1}
\end{align} 
where $P_1$ is a point on the boundary of AdS. 

\begin{figure}[t]
\centering
\includegraphics[clip,height=4.1cm]{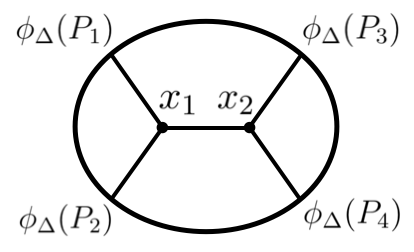}
\caption{A tree-level 4-point Witten diagram in the $\phi^3$ theory of Eq.~\ref{eq:bksd8d}. The points $x_{1,2}$ are points in the bulk, and $P_i$ are points on the boundary of AdS.}
\label{fig:12321}
\end{figure}

\begin{figure}[t]
\centering
\includegraphics[clip,height=3.6cm]{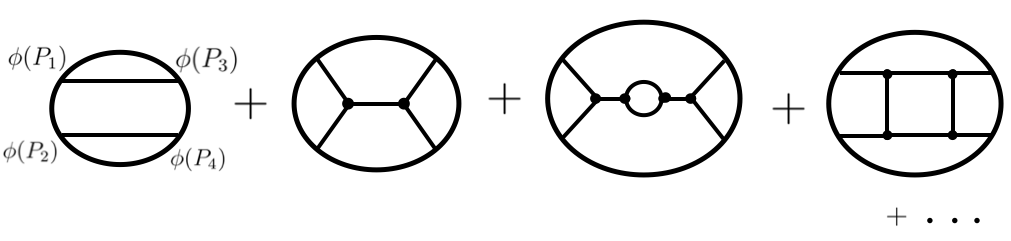}
\caption{The first few diagrams (1-loop) in the perturbative expansion of the 4-point correlator of $\phi^3$ theory in AdS. }
\label{fig:2122}
\end{figure}

Scattering amplitudes in AdS can be obtained perturbatively via the computation of Feynman diagrams in AdS. We show the tree-level 4-point Witten diagram in Fig.~\ref{fig:12321}, for the case of the $\phi^3$ theory considered in Eq.~\ref{eq:bksd8d}. The expression for this diagram is a product of four $K_\D$'s and one $G_\D$'s, which is then integrated in the bulk points $x_1$ and $x_2$ over the AdS space:
\begin{align} 
\label{eq:jhg43b}
&\langle \mm{O}_1(P_1) \ldots \mm{O}_4 (P_4) \rangle =
\nn
& \l^2 \int d^{d+1}x_1 d^{d+1}x_2 G_\D(x_1,x_2) K_{\D}(P_1,x_1)   K_{\D}(P_2,x_1)   K_{\D}(P_3,x_2)   K_{\D}(P_4,x_2)  
\end{align} 
In Fig.~\ref{fig:2122} we show the perturbative expansion of the 4-point correlator for the case of $\phi^3$ theory on AdS.

The boundary of AdS is obtained by taking the radial coordinate $\r \to \frac{\pi}{2}$. The bulk field $\phi$ is dual to the boundary operator $\mm O$:
\begin{align} 
\mm{O}(t,\O) \equiv \lim_{\epsilon \to 0} \frac{\phi(t,\r(\epsilon),\O)}{\epsilon^\D} \ \ \ \ \ \ \ , \ \ \ \ \ \ \ \ \r(\epsilon) = \frac{\pi}{2} -\epsilon e^{-\tau}
\end{align} 
The boundary operator $\mm O$ inherits its conformal transformation properties from the AdS isometries of the bulk field $\phi$, and the correlation functions of $\mm O$'s will be governed by conformal symmetry. In particular, the space-time dependence of the 2-point and 3-point correlators is entirely fixed by the conformal symmetry. The 4-point correlators will have a non-trivial space-time dependence-
\begin{align} 
\langle \mm{O}(x_1) \mm{O}(x_2) \mm{O}(x_3) \mm{O}(x_4) \sim \mm{G}(z,\bar z)
\end{align} 
where $z$ and $\bar z$ are cross-ratios built from the points $x_i$, $i=1, \ldots ,4$. We are interested in the function $\mm{G}(z,\bar z)$, which obeys an expansion in conformal blocks $\mm{K}_{J,\D}(z\bar z)$:
\begin{align} 
\mm{G}(z,\bar z) = \sum_{J,\D} c^2_{J,\D} \mm{K}_{J,\D}(z\bar z)
\end{align} 
where $\D$ and $J$ are the scaling dimensions and spin of the exchanged operators, and $c_{J,\D}$ are called OPE coefficients. A conformal theory is entirely determined from knowledge of its ``CFT data", i.e all of the scaling dimensions and OPE coefficients in the theory.

\section{The Spectral representation in AdS}
\label{sec:se1}

In this section we discuss the spectral representation of bulk 2-point functions. A bulk 2-point correlator is such that both points are inside the bulk (as opposed to being on the boundary of AdS).  We first review the spectral representation for a bulk scalar field, and then we discuss more generally the case of a spin-$J$ field. In subsection~\ref{subsec:13} we compute the 1-loop bubble 2-point function of a spin-$J$ field. In subsection~\ref{subsec:14} we attach bulk-to-boundary propagators in order to construct boundary 4-point ``blob diagrams" (to be defined below). More details about spinning AdS propagators and spectral representations can be found in \cite{Costa:2014kfa}.

\subsection{Spin-0 (scalar) spectral representation}

\begin{figure}[t]
\centering
\includegraphics[clip,height=3.5cm]{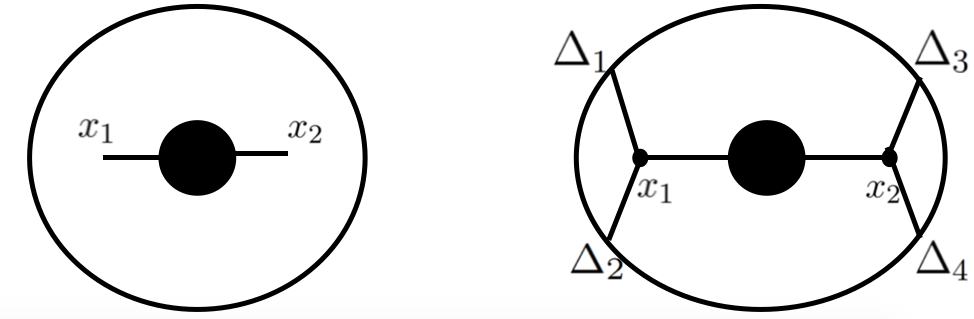}
\caption{Scalar "Blob"diagrams. \textbf{Left:} The black blob represents $F(x_1,x_2)$, a general bulk scalar two-point function. Other than that, the blob is a completely general function of the geodesic distance. \textbf{Right:} We attach 4 external scalar legs to the boundary, this gives a CFT 4-point function. This is a class of diagrams which has two bulk-to-boundary propagators attached to $x_2$, and two attached to $x_1$.}
\label{fig:bubblerelations16}
\end{figure}

Consider a scalar field $\phi(x)$ of mass $m$ in $AdS_{d+1}$ space-time. On the boundary of AdS, it corresponds to a scalar operator of scaling dimension $\D$, where $m^2= \D(\D-d)$, and we are setting the AdS radius to be $R_{AdS}=1$.  
Consider the harmonic function in $AdS_{d+1}$, which is given by the difference of bulk propagators:
\begin{align} 
\label{eq:W2}
&\O_\n (x_1,x_2) =  \frac{i\n}{2\pi} \Big( G_{\frac{d}{2}+i\n}(x_1,x_2) -G_{\frac{d}{2}-i\n}(x_1,x_2) \Big)
\end{align} 
The harmonic function is the solution of the eigenvalue equation
\begin{align} 
\pa^2_{x_1} \O_\n(x_1,x_2) = \Big( \frac{d^2}{4} +\n^2 \Big) \O_\n(x_1,x_2)
\end{align} 
The integral of $\O_\n(x_1,x_2)$ gives the delta function:
\begin{align} 
\d^{d+1}(x_1,x_2) =\int_{-\infty}^\infty d\n\ \O_\n(x_1,x_2)
\end{align} 
Any bulk 2-point function $F(x_1,x_2)$ in AdS, Fig.~\ref{fig:bubblerelations16}-left, can be expanded in a basis of AdS harmonic functions $\Omega _{\nu}(x_1,x_2)$:
\begin{align} 
\label{eq:nseeeeb}
g^{(2)}(\z)=F(x_1,x_2)= \langle \phi(x_1) \phi(x_2) \rangle= \int_{-\infty}^\infty d\n  \tilde{F}_\n  \Omega _{\nu}(x_1,x_2) \,,
\end{align}
$ \tilde{F}_\n$ is a function of $\n$, and is the spectral transform of $F(x_1,x_2)$. We wrote the LHS such as to explicitly show the dependence on the squared chordal distance. In Fig.~\ref{fig:bubblerelations16}-left we denoted the general bulk 2-point function $F(x_1,x_2)$ as a 2-point blob. The spectral representation (i.e expansion in terms of $\Omega _{\nu}(x_1,x_2)$) will be an important tool for us.

The simplest example of a bulk 2-point function is the bulk-to-bulk propagator $G_\D(x_1,x_2)=  \langle \phi(x_1) \phi(x_2) \rangle \Big|_{tree}$. We write once more the expression for the scalar bulk-to-bulk propagator:
\begin{align} 
\label{eq:fkjdf3}
G_\D(x_1,x_2) =  \frac{\G_\D}{2\pi^{\frac{d}{2}} \G_{\D-\frac{d}{2}+1}} \zeta^{-\D} {}_2F_1 (\D,\D-\frac{d-1}{2},2\D-d+1,-4\zeta^{-1})
\end{align} 
Here  $\zeta= \frac{(z_1-z_2)^2+(\vec{x}_1-\vec{x}_2)^2}{z_1z_2}$ is the chordal distance squared. This propagator can be written in the spectral representation as follows:
\begin{align} 
\label{eq:W1}
G_\D (x_1,x_2) = \int_{-\infty}^\infty \frac{ d\nu}{\n^2+(\D-\frac{d}{2})^2}  \Omega _{\nu}(x_1,x_2)  
\end{align} 
In other words, the spectral transform of the bulk-to-bulk propagator is just a simple pole $ \tilde{F}_\n =\frac{1}{\n^2+(\D-\frac{d}{2})^2}$. This can easily be seen from Eq.~\ref{eq:W2}, by using the residue theorem.

Now that we wrote the bulk 2-point function, one can obtain CFT boundary correlators by attaching bulk-to-boundary propagators $K_{\D_i}$ to the 2-point blob $F(x_1,x_2)$. As for Feynman diagrams in flat space, for each vertex one integrates over the whole AdS space. For example the 4-point correlation function (see Fig.~\ref{fig:bubblerelations16}-right) is obtained by attaching 4 bulk-to-boundary propagators:
\begin{align} 
\label{eq:jhg43}
\langle \mm{O}_1 \ldots \mm{O}_4  \rangle =
  \int d^{d+1}x_1 d^{d+1}x_2 F(x_1,x_2) K_{\D_1}(P_1,x_1)   K_{\D_2}(P_2,x_1)   K_{\D_3}(P_3,x_2)   K_{\D_4}(P_4,x_2)  
\end{align} 
Here the $P_i$ are boundary points, and $x_i$ are bulk points. We denote the family of AdS Diagrams in Fig.~\ref{fig:bubblerelations16}-right as ``Blob diagrams".

\subsection{Spin-$J$ spectral representation}

\begin{figure}[t]
\centering
\includegraphics[clip,height=3.5cm]{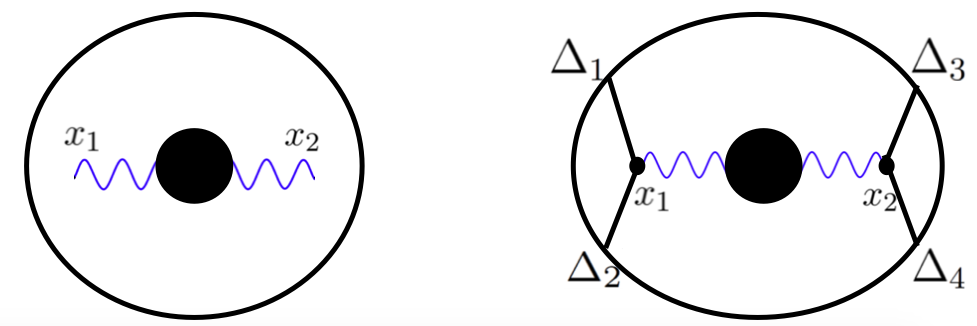}
\caption{Spinning ``Blob"diagrams. \textbf{Left:} The black blob with wavy lines represents a general bulk spin-$J$ two-point function. Other than that, the blob is completely general. \textbf{Right:} We attach 4 external scalar legs to the boundary, this gives a CFT 4-point function. This is a class of diagrams which has two bulk-to-boundary propagators attached to $x_2$, and two attached to $x_1$.}
\label{fig:bubblerelations161}
\end{figure}

The results of the previous section can be generalised to include spinning fields.
We denote the bulk-to-bulk propagator of a spin-$J$ field in AdS by $G_{\D,J}(x_1,x_2,W_1,W_2)$, where $W_{1,2}$ are polarisation vectors. The bulk propagator of the spin-$J$ field has a spectral representation \cite{Costa:2014kfa}: 
\begin{align} 
G_{\D,J}(x_1,x_2,W_1,W_2) = \sum_{l=0}^{J} ((W_1 \cdot \nabla_1) (W_2 \cdot \nabla_2))^{J-l} \int_{-\infty}^\infty d\n a_l(\n) \O_{\n,l}(x_1,x_2:W_1,W_2)
\nn
\end{align} 
where $\O_{\n,l}$ is the AdS harmonic function of spin-$l$:
\begin{align} 
\O_{\n,l} (x_1,x_2,W_1,W_2) = \frac{i \n}{2 \pi } \Big(  G_{\frac{d}{2}+i\n,l} (x_1,x_2,W_1,W_2) -G_{\frac{d}{2}-i\n,l} (x_1,x_2,W_1,W_2) \Big)
\nn
\end{align} 
One can easily find the coefficients $a_l(\n)$, for example for $l=J$ we have $a_J(\n) = \frac{1}{\n^2+(\D-\frac{d}{2})^2}$, see \cite{Costa:2014kfa}.

For a general bulk 2-point function $F_J(x_1,x_2,W_1,W_2)$ of spin-$J$, we have the following spectral representation:
\begin{align} 
\label{eq:94n3}
F_J(x_1,x_2,W_1,W_2) = \sum_{l=0}^{J} ((W_1 \cdot \nabla_1) (W_2 \cdot \nabla_2))^{J-l} \int_{-\infty}^\infty d\n F_{\n,l}\O_{\n,l}(x_1,x_2:W_1,W_2)
\nn
\end{align} 
Where the harmonic function in AdS is:
\begin{align} 
\O_{\n,l} (x_1,x_2,W_1,W_2) = \frac{i \n}{2 \pi } \Big(  G_{\frac{d}{2}+i\n,l} (x_1,x_2,W_1,W_2) -G_{\frac{d}{2}-i\n,l} (x_1,x_2,W_1,W_2) \Big)
\nn
\end{align} 
which generalises Eq.~\ref{eq:W2}. The harmonic function has a split representation \cite{Costa:2014kfa}:
\begin{align} 
\O_{\n,l} (x_1,x_2,W_1,W_2) = \frac{\n^2}{\pi l! (\frac{d}{2}-1)_l} \int d^dP K_{\frac{d}{2}+i\n,l} (x_1,P,W_1,D_Z) K_{\frac{d}{2}-i\n,l} (x_2,P,W_2,Z)
\nn
\end{align} 
where $K_{\frac{d}{2}+i\n,l}$ is the spin-$l$ bulk-to-boundary propagator, and the integration is over the boundary of AdS.

\subsection{The 1-loop bubble spectral representation}
\label{subsec:13}

\subsubsection{Scalar field case}
In the previous section we saw the spectral representation of the bulk propagator. Here we look at the spectral representation of 1-loop bubble, which is just a product of two bulk propagators $B(x,y) \equiv \langle \phi^2(x) \phi^2(y)\rangle =G^2_\D(x,y)$. We write the spectral representation \cite{Carmi:2018qzm}:
\begin{align} 
B(x,y) = \int_{-\infty}^\infty d\n \tilde{B}(\n)  \O_\n(x,y)
\end{align} 
In \cite{Carmi:2018qzm} we computed the spectral transform of bubble $\tilde{B}(\n)$.  For the case of $AdS_3$, it takes the simple form:
\begin{align} 
\tilde{B}(\n) = \frac{i}{8\pi \n} \Big[ \psi(\D-\frac{1+i\n}{2}) -\psi(\D-\frac{1-i\n}{2}) \Big]
\end{align} 
where $\psi$ is the digamma function $\psi(z) \equiv \frac{d}{dz} \log[\G(z)]$, and $\D$ is the scaling dimension of the scalar field. This expression further simplifies when $\D=1$:
\begin{align} 
\label{eq:true467}
B(\n)\Big|_{\D=1} = -\frac{1}{8 i\n}\cot (\frac{\pi}{2}(1+i\n)) =\frac{1}{16} \frac{\G_{\frac{1}{2}+\frac{i \n}{2} } \G_{\frac{1}{2}-\frac{i \n}{2} } }{\G_{1- \frac{i \n}{2} } \G_{1+ \frac{i \n}{2} }  }
\end{align}

\subsubsection{The $U(1)$ current case}
In \cite{Ankur:2023lum} we studied large-$N_f$ scalar quantum electrodynamics on $AdS_{d+1}$. The two-point function of the $U(1)$ conserved currents has the spectral representation:
 \begin{align} 
\langle j_M(x) j_N(y)  \rangle = - \int_{-\infty}^\infty d\n B^{(1)}(\n) \O^{(1)}_{\n M N}(x,y)
\end{align} 
where $B^{(1)}(\n)$ is the 1-loop bubble in the spectral representation, and $\O^{(1)}_{\n M N}(x,y)$ is the spin-1 AdS harmonic function. 
For a massless gauge field in $AdS_3$, we derived \cite{Ankur:2023lum}:
\begin{align} 
B^{(1)}(\n) = \frac{\n}{16\pi (\n^2+1)} \Big( -2(2\D-3)\n +(\n^2 +4(\D-1)^2)(i\psi(\D-\frac{i \n}{2})- i\psi(\D+\frac{i \n}{2})) \Big)
\nn
\end{align}
where $\D$ is the scaling dimension of the scalars. This expression simplifies further when $\D=1$:
\begin{align} 
\label{eq:true4}
B^{(1)}(\n)\Big|_{\D=1} = \frac{\n^3}{16(\n^2+1)} \coth (\frac{\pi \n}{2}) =  -\frac{i \n^3}{16(\n^2+1)} \frac{\G_{1+\frac{i \n}{2} } \G_{-\frac{i \n}{2} } }{\G_{\frac{1}{2}- \frac{i \n}{2} } \G_{\frac{1}{2}+ \frac{i \n}{2} }  }
\end{align} 
We will use these results in the following to derive identities for AdS diagrams.

\subsection{Boundary correlators: Blob diagrams}
\label{subsec:14}

One constructs boundary correlators by attaching bulk-to-boundary propagators to the bulk 2-point blob, Fig.~\ref{fig:bubblerelations161}. We previously saw this for the case of scalars, around Eq.~\ref{eq:jhg43}, and here we show the generalisation for a spin-J 2-point blob. For instance the 4-point correlators blob diagrams \cite{Costa:2014kfa}:
\begin{align} 
\label{eq:nckd89}
&\langle \mm{O}_1(P_1) \ldots \mm{O}_4(P_4)  \rangle =
\int d^{d+1}x_1 d^{d+1}x_2 F_J(x_1,x_2) \times
\nn
&K_{\D_1}(P_1,x_1) [ (K_1\cdot \nabla_1)^J  K_{\D_2}(P_2,x_1) ] K_{\D_3}(P_3,x_2) {}[(K_2\cdot \nabla_2)^J  K_{\D_4}(P_4,x_2) ] 
\end{align} 
The spectral representation of $F_J(x_1,x_2)$ is given in Eq.~\ref{eq:94n3}. One can combine Eqs.~\ref{eq:94n3} and \ref{eq:nckd89} to get \cite{Costa:2014kfa}:
\begin{align} 
\label{eq:dfvnjg}
\langle \mm{O}_1(P_1) \ldots \mm{O}_4(P_4)  \rangle = 
A \sum_{l=0}^J \int_{-\infty}^\infty d\n F_{\n,l}   \mathcal{P}^{\D_i}_{\frac{d}{2}+i\nu,l} (z,\bar z)
\end{align} 
The RHS is the conformal partial wave expansion of the 4-point correlator on the LHS.
The RHS is written in terms of conformal cross-ratios $u =z\bar z = \frac{P_{12}P_{34}}{P_{13}P_{24}}$ and $v=(1-z)(1-\bar z)= \frac{P_{14}P_{23}}{P_{13}P_{24}} $. For notational simplicity, we defined:
\begin{align} 
\label{eq:nvjdf8}
A\equiv 2 g_{\phi_1\phi_2h}g_{\phi_3\phi_4h} \frac{  \Big( \frac{P_{24} }{P_{14}} \Big)^{a}  \Big( \frac{P_{14} }{P_{13}} \Big)^{b}  }{(P_{12})^{a_p }(P_{34})^{b_p}}
\end{align} 
Likewise we defined $a\equiv \frac{\D_1-\D_2}{2}$, $b\equiv \frac{\D_3-\D_4}{2}$ and $a_p\equiv \frac{\D_1+\D_2}{2}$, $b_p\equiv \frac{\D_3+\D_4}{2}$. In Eq.~\ref{eq:dfvnjg}, $\mathcal{P}^{\D_i}_{\frac{d}{2}+i\nu,l} (z,\bar z)$ is the conformal partial wave, which can be obtained via a convolution of 3-point functions:
\begin{align}
\mathcal{P}^{\D_i}_{\frac{d}{2}+i\nu,l} (z,\bar z) = \int dP \langle \mm{O}_{\D_1}(P_1) \mm{O}_{\D_2}(P_2) \mm{O}_{\frac{d}{2}+i\n,l}(P,D_z) \rangle  \langle \mm{O}_{\frac{d}{2}-i\n,l}(P,Z) \mm{O}_{\D_3}(P_3) \mm{O}_{\D_4}(P_4)  \rangle 
\end{align}
The integration above is over the boundary of AdS. Defining
\begin{align} 
\label{eq:dfjhf9f}
 g_l(z,\bar z) \equiv \frac{1}{\pi^2} \int_{-\infty}^\infty d\n F_{\n,l}   \mathcal{P}^{\D_i}_{\frac{d}{2}+i\nu,l} (z,\bar z) \ \ \ \ \ , \ \ \ \ \ 
\end{align} 
Eq.~\ref{eq:dfvnjg} becomes-
\begin{align} 
\label{eq:sal23}
\langle \mm{O}_1(P_1) \ldots \mm{O}_4(P_4)  \rangle \equiv  A g^{(4)}(z, \bar z)  = A \pi^2  \sum_{l=0}^J g_l(z,\bar z)
\end{align}

Now we want to write Eq.~\ref{eq:dfjhf9f} in terms of conformal blocks $\mathcal{K}^{\D_i}_{\frac{d}{2}+i\nu,l} (z,\bar z)$ instead of conformal partial waves $\mathcal{P}^{\D_i}_{\frac{d}{2}+i\nu,l} (z,\bar z)$. Recall that $\mathcal{P}^{\D_i}_{\frac{d}{2}+i\nu,l} (z,\bar z)= C\times \mathcal{K}^{\D_i}_{\frac{d}{2}+i\nu,l} (z,\bar z)$, where $C$= products of gamma functions. Using this fact, we get that\footnote{In Eq.~\ref{eq:98d3} we have defined \begin{align} 
\label{eq:sdskkf}
\Upsilon_{\nu,l}^{\D_i} \equiv \frac{\Bigg[ \frac{  \G_{a_p+ \frac{l}{2} -\frac{d}{4}+\frac{i \nu}{2}}  \G_{b_p+\frac{l}{2} -\frac{d}{4}+\frac{i \nu}{2}}  }{\G_{1-a_p-\frac{l}{2} +\frac{d}{4}+\frac{i \nu}{2}}  \G_{1-b_p-\frac{l}{2} +\frac{d}{4}+\frac{i \nu}{2}}  }  \Bigg] \Big( \G_{a+\frac{l+i \nu+\frac{d}{2}}{2}} \G_{-a+\frac{l+i \nu+\frac{d}{2}}{2}} \G_{b+\frac{l+i \nu+\frac{d}{2}}{2}} \G_{-b+\frac{l+i \nu+\frac{d}{2}}{2}} \Big) }{\Gamma_{i\nu} \Gamma_{\frac{d}{2}+i\n+l}  (i\n + \frac{d}{2}-1)_l  } \ \ \ . 
\end{align} where $ (i\n + \frac{d}{2}-1)_l$ is the Pochhammer symbol.}
\begin{align} 
\label{eq:98d3}
\boxed{ g_l(z,\bar z)  = \int_{-\infty}^\infty d\nu 
\frac{ \tilde{F}_{\n,l}  }{  \sin \pi (a_p-\frac{\frac{d}{2}+i \nu-l}{2}) \sin \pi (b_p-\frac{\frac{d}{2}+i \nu-l}{2})}  
 \Upsilon_{\nu,l}^{\D_i} \times \mathcal{K}^{\D_i}_{\frac{d}{2}+i\nu,l} (z,\bar z) }
\end{align}
Eq.~\ref{eq:sal23} and \ref{eq:98d3} give the expansion of the 4-point correlator in terms of conformal blocks $\mathcal{K}^{\D_i}_{\frac{d}{2}+i\nu,l} (z,\bar z)$. The poles from the denominator $\sin \pi (a_p-\frac{\frac{d}{2}+i \nu-l}{2}) \sin \pi (b_p-\frac{\frac{d}{2}+i \nu-l}{2})$ give the contribution of the double-trace operators to the conformal block expansion. These poles are located at $\frac{d}{2}+i \nu = 2a_p+2n+l$ and $\frac{d}{2}+i \nu = 2b_p+2n+l$, where $n$ is an integer. Additional poles will come from the function $ \tilde{F}_{\n,l}$.


Now that we wrote the 4-point function $\mathcal{P}^{\D_i}_{\frac{d}{2}+i\nu,l} (z,\bar z)$ in Eq.~\ref{eq:sal23}, we can take the double-discontinuity of it, \cite{Caron-Huot:2017vep}. Taking the double discontinuity simply cancels the double-trace poles arising from the $sin$ denominators in Eq.~\ref{eq:98d3}:
\begin{align} 
\label{eq:fdf33}
dDisc[ g_l(z, \bar z)]=    \int_{-\infty}^\infty d\nu 
 \tilde{F}_{\n,l}  \Upsilon_{\nu,l}^{\D_i} \times \mathcal{K}^{\D_i}_{\frac{d}{2}+i\nu,l} (z,\bar z) 
\end{align} 
One can compute the RHS above by using the residue theorem:
\begin{align} 
\label{eq:df54}
dDisc[ g_l(z, \bar z)] =   \sum_{n=0}^\infty 
Res[\tilde{F}_{\n,l}]  \Upsilon_{\nu,l}^{\D_i} \times \mathcal{K}^{\D_i}_{\frac{d}{2}+i\nu,l} (z,\bar z) \Big|_{\n=\n_n}
\end{align} 
where $\n_n$ are the poles of $F_{\n,l}$ and $ Res[\tilde{F}_{\n,l}]$ are their residues.

Alternatively, we can take a single discontinuity, which results in cancelling one of the two do $\sin$ factors in Eq.~\ref{eq:98d3}: 
\begin{align} 
\label{eq:dfjkhv8}
Disc_{a_p} [g_l(z, \bar z)] =    \int_{-\infty}^\infty d\nu 
\frac{ \tilde{F}_{\n,l}  }{  \sin \pi (b_p-\frac{\frac{d}{2}+i \nu-l}{2})}  
 \Upsilon_{\nu,l}^{\D_i} \times \mathcal{K}^{\D_i}_{\frac{d}{2}+i\nu,l} (z,\bar z) 
\end{align} 
For example, the simplest 4-point blob diagram is the tree-level contact diagram, for which $\tilde{F}_{\n,l} =1$. The previous expression Eq.~\ref{eq:dfjkhv8} can then be computed using the residue theorem:
\begin{align} 
\label{eq:dfkss}
Disc_{a_p} [g_l(z, \bar z)] =   \sum_{n=0}^\infty  \Upsilon_{\nu,l}^{\D_i} \times \mathcal{K}^{\D_i}_{\frac{d}{2}+i\nu,l} (z,\bar z) \Big|_{\frac{d}{2}+i\nu \to b_p+l+n}
\end{align} 
Let us make a few more convenient definitions regarding the gamma functions appearing in Eq.~\ref{eq:sdskkf}:
\begin{align} 
\label{eq:ghsdkj3}
U_{A.V}^{(a_p,l)} \equiv \G_{a_p+\frac{l}{2} +\frac{i \nu-\frac{d}{2}}{2}} \G_{a_p+\frac{l}{2} -\frac{i \nu+\frac{d}{2}}{2}} 
\end{align} 
and
\begin{align} 
\label{eq:ghsdkj4}
&U_{V}^{(a_p,l,\D)}  \equiv \G_{a_p+\frac{l}{2} +\frac{i \nu-\frac{d}{2}}{2}} \G_{a_p+\frac{l}{2} -\frac{i \nu+\frac{d}{2}}{2}} \times \frac{1}{\n^2+(\D-\frac{d}{2})}
\end{align}

\section{Identities for Witten diagrams}
\label{sec:se2}

In this section we derive various vertex/propagator identities, which will enable to relate different Witten diagrams together. These identities are generalisations and extensions of the identities that we derived in \cite{Carmi:2019ocp,Carmi:2021dsn}

\subsection{Identity 0}
\label{sec:dmmd}

For a $CFT_d$ with $d=$even space-time dimensions. The 4-point conformal blocks contain the following hypergeometric function:
\begin{align} 
k_\b^{(a,b)}(z) \equiv z^{\frac{\b}{2}} {}_2 F_1 (\frac{\b}{2}+a,\frac{\b}{2}+b,\b,z)
\end{align}  
Recall that the conformal blocks in $d=2$ and $d=4$ are:
\begin{align} 
\mm{K}^{(d=2)}_{\D,l} (z,\bar z)  = \frac{1}{1+\d_{l,0}} \Big(k_{\D-l}^{(a,b)}(z) k_{\D+l}^{(a,b)}(\bar z) + k_{\D+l}^{(a,b)}(z) k_{\D-l}^{(a,b)}(\bar z) \Big)
\nn
\mm{K}^{(d=4)}_{\D,l} (z,\bar z)  = \frac{z\bar z}{z-\bar z} \Big(k_{\D-l-2}^{(a,b)}(z) k_{\D+l}^{(a,b)}(\bar z) - k_{\D+l}^{(a,b)}(z) k_{\D-l-2}^{(a,b)}(\bar z) \Big)
\end{align} 
Following \cite{Dolan:2011dv}, lets define the following differential operators:
\begin{align} 
&D^z_{+} \equiv z^{1-b} \pa_ z z^b  
\nn
&D^z_{-}f(z) \equiv z^{1+b} (1-z)^{1-a-b} \pa_z \Big[ z^{-b} (1-z)^{a+b}f(z)\Big]
\end{align} 
Acting with them on $k_\b^{(a,b)}(z)$, raises or lowers $b$ by one \cite{Dolan:2011dv}:
\begin{align} 
&D^z_{+} k_\b^{(a,b)}(z) = (\frac{\b}{2}+b) k_{\b}^{(a,b+1)}(z) 
\nn
&D^z_{-} k_\b^{(a,b)}(z) = (\frac{\b}{2}-b) k_{\b}^{(a,b-1)}(z) 
\end{align} 
Since the functions $k_\b^{(a,b)}(z)$ are symmetric with respect to to $a$ and $b$, one can likewise define raising/lowering operators for $a$.

\subsubsection{$d=2$}
In $d=2$, the spectral representation of the 4-point function, Eq.~\ref{eq:sdskkf} and \ref{eq:98d3}, contains the following factor $ \G_{\frac{i\n+l +1}{2}+b} \G_{\frac{i\n+l +1}{2}-b} k_{\frac{i\n+l +1}{2}}^{(a,b)}(z) k_{\frac{i\n -l+1}{2}}^{(a,b)}(\bar z)$. Acting with two raising operators gives: 
\begin{align} 
&D^{\bar z}_{+} D^z_{+} \Big[ \G_{\frac{i\n+l +1}{2}+b} \G_{\frac{i\n+l +1}{2}-b} k_{i\n+l +1}^{(a,b)}(z) k_{i\n -l+1}^{(a,b)}(\bar z) \Big] 
\nn
&=\frac{-1}{4} (\n^2+( -l+2b+1)^2)  \Big(\G_{\frac{i\n +l+1}{2}+b+1}  \G_{\frac{i\n+l +1}{2}-b-1}  k_{i\n +1+l}^{(a,b+1)}(z)  k_{i\n -l+1}^{(a,b+1)}(\bar z) \Big)
\end{align} 
This factor $(\n^2+( -l+2b+1)^2)$ is the inverse propagator with $\d=2b+2-l$, see \ref{eq:W1}. Therefore:
\begin{align} 
\label{eq:nbvl3}
 \textbf{Identity\ 0a:} \ \ \ \ \ \ \ \ \boxed{D^{\bar z}_{+} D^z_{+}  g^{(a,b)}_l(z,\bar z)  = \frac{-1}{4}  (\n^2+( -l+2b+1)^2) g^{(a,b+1)}_l (z,\bar z) }
\end{align} 
This is shown in Fig.~\ref{fig:bubblerelations1625}.

\subsubsection{$d=4$}
\begin{figure}[t]
\centering
\includegraphics[clip,height=3.5cm]{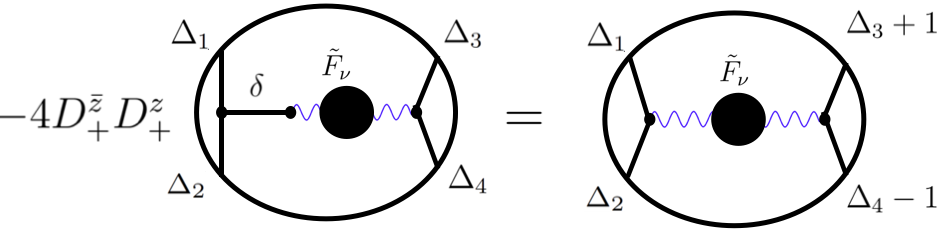}
\caption{Showing identity 0 in Eqs.~\ref{eq:nbvl3} and \ref{eq:nbvl4}. In the LHS we have a scalar propagator with scaling dimension  $\d =2b-l+2$.}
\label{fig:bubblerelations1625}
\end{figure}

For $d=4$:
\begin{align} 
&D^{\bar z}_{+} D^z_{+} \Big[ \G_{\frac{i\n+l +2}{2}+b} \G_{\frac{i\n+l +2}{2}-b} k_{i\n+l+2}^{(a,b)}(z) k_{i\n -l}^{(a,b)}(\bar z) \Big]
\nn
& = \frac{-1}{4} (\n^2+( -l+2b)^2) \Big[ \G_{\frac{i\n+l +2}{2}+b+1} \G_{\frac{i\n+l +2}{2}-b-1} k_{i\n+l+2}^{(a,b+1)}(z) k_{i\n -l}^{(a,b+1)}(\bar z) \Big]
\end{align} 
Therefore
\begin{align} 
\label{eq:nbvl4}
 \textbf{Identity\ 0b:} \ \ \ \ \ \ \ \ \boxed{D^{\bar z}_{+} D^z_{+}  g^{(a,b)}_l(z,\bar z)  = \frac{-1}{4}  (\n^2+( 2b-l)^2) g^{(a,b+1)}_l (z,\bar z) }
\end{align} 
The factor is is the inverse propagator with $\d =2b-l+2$. This identity is shown in Fig.~\ref{fig:bubblerelations1625}.

\subsection{Identity 1}

Looking at  Eq.~\ref{eq:98d3}, $g_l^{(\D_i')}(z,\bar z)$ obeys the following identity:
\begin{align} 
 \textbf{Identity\ 1:} \ \ \ \ \ \ \ \ \ \ \ \  \boxed{  g_l^{(\D_i')}(z,\bar z) =   g_l^{(\D_i)} (z, \bar z) }
\end{align} 
where $\D_i= (\D_1,\D_2,\D_3,\D_4)$ and $\D_i'$ is 
\begin{align} 
\label{eq:sdbsmd7s}
\D_1'= \D_{134,2} \ \ \ ,\ \ \ \D_2'= \D_{234,1} \ \ \ ,\ \ \ \D_3'= \D_{123,4} \ \ \ ,\ \ \ \D_4'= \D_{124,3}
\end{align} 
and we defined $\D_{123,4}\equiv \frac{\D_1+\D_2+\D_3-\D_4}{2}$. This identity works exactly the same as for scalar vertex, so it is a simple generalization of identity 1 in section 3.1 of \cite{Carmi:2019ocp}.
 
\subsection{Identity 2}

The conformal block obeys the following eigenvalue equation:
\begin{align} 
\label{eq:dfdd}
&2D_{z,\bar z} \mm{K}_{\frac{d}{2}+i\n,l} = - \Big[ \n^2 +\frac{d^2}{4} -l(d+l-2) \Big]   \mm{K}_{\frac{d}{2}+i\n,l}
\end{align} 
where the above differential operator is defined as:
\begin{align} 
\label{eq:dfdd23}
&D_{z,\bar z} \equiv [z^2(1-z)\pa_z^2-(a+b+1)z^2\pa_z-abz+ z \leftrightarrow \bar z]
\nn
&+(d-2)\frac{z\bar z}{z-\bar z} [(1-z)\pa_z-(1-\bar z)\pa_{\bar z}]
\end{align} 
We define (see Eq.~\ref{eq:ghsdkj3}):
\begin{align} 
\label{eq:vmnsd9}
U_{A.V}^{(a_p,l)} \equiv \G_{a_p+\frac{l}{2} +\frac{i \nu-\frac{d}{2}}{2}} \G_{a_p+\frac{l}{2} -\frac{i \nu+\frac{d}{2}}{2}} 
\end{align} 
thus
\begin{align}
U_{A.V}^{(a_p-1,l)}\times \Big(\frac{-1}{4}\Big)\Big[ 2D_{z,\bar z} -\a)\Big] \mm{K}_{\frac{d}{2}+i\n,l} =U_{A.V}^{(a_p,l)}  \mm{K}_{\frac{d}{2}+i\n,l}
\end{align}
where we defined:
\begin{align}
\a \equiv 4 a_p^2 -2a_p (d+4-2l)+2(d+1+l(l-3)
\end{align}
In other words
\begin{align}
U_{A.V}^{(a_p-1,l)} \mm{D}^{(+)}_{z,\bar z} \mm{K}_{\frac{d}{2}+i\n,l} =U_{A.V}^{(a_p,l)}  \mm{K}_{\frac{d}{2}+i\n,l} 
\end{align}
where we defined the differential operator:
\begin{align}
\mm{D}^{(+)}_{z,\bar z}  \equiv \Big(-\frac{1}{4}\Big)\Big[ 2D_{z,\bar z} - \a\Big] 
\end{align}
We can write this more neatly in terms of the 4-point function $g_l$ of Eq.~\ref{eq:98d3}, which gives 
\begin{align} 
\label{eq:relation22}
\textbf{Identity\ 2:}  \ \ \ \ \ \ \boxed{g_l^{(\D_1,\D_2,\D_3,\D_4)}(z, \bar z) = \mm{D}_{z,\bar z}^{(+)} \ \ g_l^{(\D_1-1,\D_2-1,\D_3,\D_4)}(z, \bar z)  }
\end{align}
This identity is a generalization of identity 2 of \cite{Carmi:2019ocp} to spinning operators. This identity enables to raise $\D_1$ and $\D_2$ by integers.
 
\subsection{Identity 3}

We define (see Eq.~\ref{eq:ghsdkj4}):
\begin{align} 
\label{eq:ghsdkj45}
&U_{V}^{(a_p,l,\D)}  \equiv \G_{a_p+\frac{l}{2} +\frac{i \nu-\frac{d}{2}}{2}} \G_{a_p+\frac{l}{2} -\frac{i \nu+\frac{d}{2}}{2}} \times \frac{1}{\n^2+(\D-\frac{d}{2})}
\end{align} 
If we choose $\D=2a_p+l-2$ and recall Eq.~\ref{eq:vmnsd9}, we get:
\begin{align} 
\label{eq:145}
 \textbf{Identity\ 3:} \ \ \ \ \ \ \ \ \boxed{U_{V}^{(a_p,\D=2a_p+l-2)}   =\frac{1}{4} U_{A.V}^{(a_p-1)}  }
\end{align} 
This identity is a generalization of identity 3 of \cite{Carmi:2019ocp} to spinning operators.

\subsection{Identity 4}

\begin{figure}[t]
\centering
\includegraphics[clip,height=3.5cm]{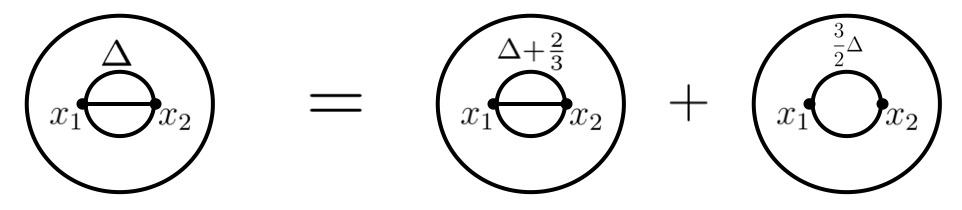}
\caption{Showing identity 4a of Eq.~\ref{eq:jnfd0d}: Relating 2-loop sunset diagrams with different scaling dimensions to the 1-loop bubble diagram.}
\label{fig:bubblerelations162}
\end{figure}

The bulk propagator squared (i.e the 1-loop bubble) can be written as an infinite sum of bulk propagators, as was derived in \cite{Fitzpatrick:2010zm,Fitzpatrick:2011hu}. In $AdS_3$ this takes the following form
\begin{align}
\label{eq:bnmdf5}
G_\D^2 (x_1,x_2)= \sum_{n=0}^\infty  G_{2\D+2n}(x_1,x_2)
\end{align}
Now let's look at the sunset diagram in AdS, which in position space is just a product of three propagators, see Fig.~\ref{fig:bubblerelations162}.
In $d=2$ dimensions, the sunset diagram gives with propagators of scaling dimension $\d$ is (see \cite{Fitzpatrick:2010zm,Fitzpatrick:2011hu}:):
\begin{align}
G_\d^3 (x_1,x_2)= \sum_{n=0}^\infty (n+1) G_{3\d+2n}(x_1,x_2)
\end{align}
Now if we plug $\d = \D+\frac{2}{3}$, we get:
\begin{align}
G_{\D+\frac{2}{3}}^3 (x_1,x_2) = \sum_{n=0}^\infty n G_{3\D+2n}(x_1,x_2)
\end{align}
Subtracting the previous two equations gives:
\begin{align}
G_{\D}^3(x_1,x_2)-G_{\D+\frac{2}{3}}^3(x_1,x_2) = \sum_{n=0}^\infty  G_{3\D+2n}(x_1,x_2)= G_{\frac{3}{2}\D}^2(x_1,x_2) =B_{\d=\frac{3}{2} \D}(x_1,x_2)
\end{align}
where in last two equalities we used Eq.~\ref{eq:bnmdf5}.
The RHS is just a bubble diagram (a bulk propagator squared). Therefore in the spectral representation the above equation becomes:
\begin{align} 
\label{eq:jnfd0d}
 \textbf{Identity\ 4a:} \ \ \ \ \ \ \ \ \ \ \ \  \boxed{ B^{\D}_{sun}(\n) - B^{\D+\frac{2}{3}}_{sun}(\n) = B^{\frac{3\D}{2}}(\n) }
\end{align}
This identity is shown in Fig.~\ref{fig:bubblerelations162}. This identity relates sunset diagrams with different scaling dimensions.

\begin{figure}[t]
\centering
\includegraphics[clip,height=2.0cm]{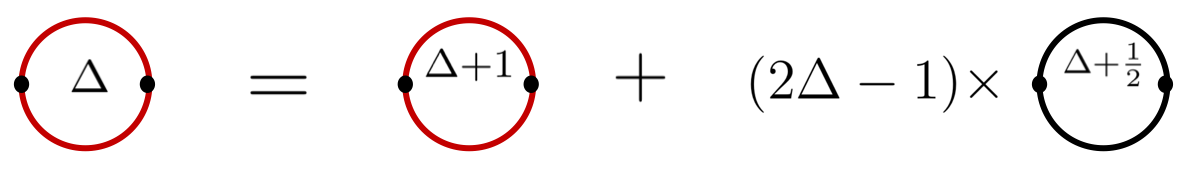}
\caption{Showing identity 4b of Eq.~\ref{eq:jnfd0d2}: Relating 1-loop fermionic bubbles. The red lines denote fermionic bulk propagators, whereas the black line denote scalar propagators. This identity can be used inside of Witten diagrams.}
\label{fig:benda1}
\end{figure}

Now let us derive a similar identity for the fermionic bubble. Consider the fermionic 1-loop bubble $(G^{F}_\D)^2$, where $G^{F}$ is the bulk propagator of fermionic field. The fermionic bubble in $AdS_3$ with scaling dimension $\D$ is (see section 6 of \cite{Carmi:2018qzm}):
\begin{align} 
(G^{F}_\D)^2 = \sum_{n=0}^\infty (n+1) (2\D+n-1)G_{2\D+2n+1}
\end{align} 
Therefore with scaling dimension $\D\to \D+1$, we have:
\begin{align} 
(G^{F}_{\D+1})^2  =\sum_{n=0}^\infty n (2\D+n)G_{2\D+2n+1}
\end{align} 
Subtracting the previous two equations gives the scalar bubble function $G^{B}$ :
\begin{align} 
(G^{F}_\D)^2-(G^{F}_{\D+1})^2 =  (2\D-1) \sum_{n=0}^\infty  G_{2\D+2n+1} =  (2\D-1) (G^{B}_{\D+\frac{1}{2}})^2
\end{align} 
This gives identity 4b, for the 1-loop bubble of fermions:
\begin{align} 
\label{eq:jnfd0d2}
 \textbf{Identity\ 4b:} \ \ \ \ \ \ \ \ \ \ \ \  \boxed{ (G^{F}_\D)^2 =  (G^{F}_{\D+1})^2 + (2\D-1) (G^{B}_{\D+\frac{1}{2}})^2 }
\end{align} 
where $G^{B}_{\D+\frac{1}{2}}$ is the scalar bubble. This identity is shown in Fig.~\ref{fig:benda1}.

\subsection{Identity 6}

\begin{figure}[t]
\centering
\includegraphics[clip,height=10.2cm]{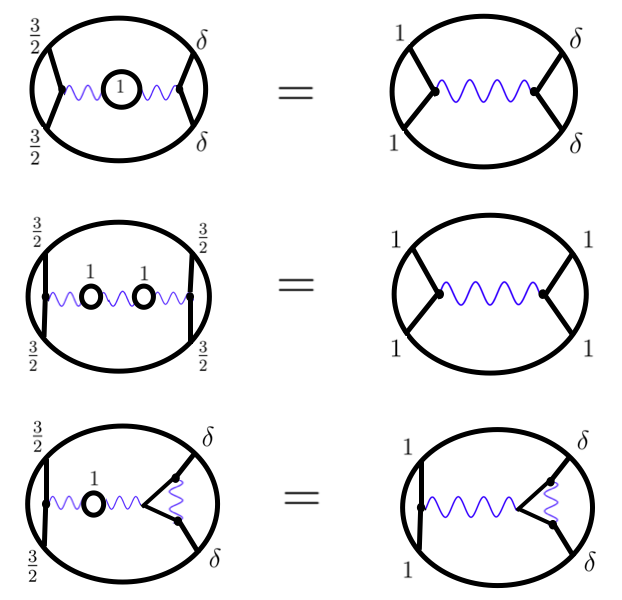}
\caption{Using identity 6 to reduce loop diagrams to lower loop diagrams in several examples. $\d$ is a general scaling dimension. Top: A 1-loop bubble diagram that reduces to the tree-level exchange diagram on the right. Middle: A 2-loop bubble diagram that reduces to the tree-level exchange diagram. Bottom: A 2-loop diagram that reduces to a 1-loop diagram.} \label{fig:bbd6d} 
\end{figure}

Let us define
\begin{align} 
\label{eq:nvd6}
U_B^{(a_p,\D,l)} \equiv \frac{1}{\n^2} \G_{ a_p+\frac{l}{2}-\frac{d}{4}-\frac{i\n }{2}} \G_{a_p+ \frac{ l}{2}-\frac{d}{4}+\frac{i\n }{2}} B^{(\D,l)}_\n
\end{align} 
where $B^{(\D,l)}_\n$ is the spectral representation of the 1-loop bubble, see section~\ref{subsec:13}. Let's focus on the case of $AdS_3$ and $l=1$ and $\D=1$ in the bubble., In this case the bubble spectral representation simplifies to a product of gamma functions, as in Eq.~\ref{eq:true4}. From Eqs.~\ref{eq:nvd6}, \ref{eq:true4}. and \ref{eq:ghsdkj3}, we get
\begin{align} 
\textbf{Identity\ 6:}  \ \ \ \ \ \ \boxed{ U_B^{(a_p=\frac{3}{2},\D=1,l=1)} =\frac{1}{32} U_{A.V}^{(a_p=1,l=1)} }
\end{align} 
This identity will help relate loop diagrams to lower loop diagrams or tree level diagrams. We show several examples of this in Fig.~\ref{fig:bbd6d}. 

\section{Large-$N_f$ scalar QED on AdS}
\label{sec:se3}

\begin{figure}[t]
\centering
\includegraphics[clip,height=4.0cm]{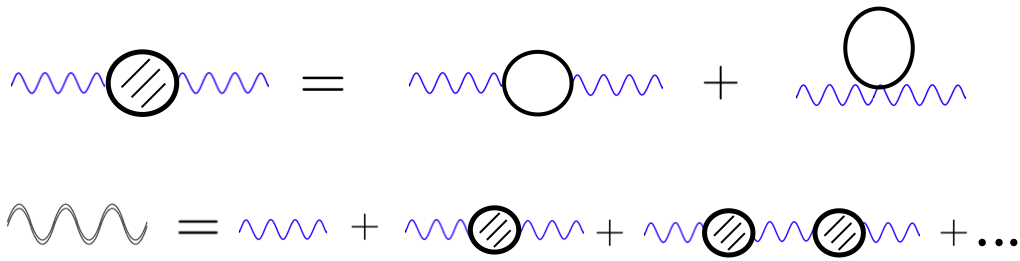}
\caption{Top: The 1PI 1-loop correction to photon propagator, which consists of 2 diagrams. Bottom: In scalar QED, the exact photon propagator at order $\frac{1}{N_f}$is given by the sum of bubble diagrams.} \label{fig:figd33438} 
\end{figure}

\subsection{Large-$N_f$ scalar QED on flat space}

Quantum electrodynamics in different dimensions is a useful arena for studying strongly coupled gauge theories \cite{Pisarski:1984dj,Appelquist:1986fd,Appelquist:1988sr,Roberts:1994dr}.  Considering the theory with $N_f$ scalar matter fields (sQED) at large $N_f$ and dimension $2<D<4$ the theory is asymptotically free and has an interesting phase structure.
Consider $N_f$ scalar fields $\phi^a$ and a gauge field $A_\m$, where the sQED Lagrangian with a quartic coupling is:
 \begin{align} 
 \label{eq:vkdfk4}
L= \frac{1}{4e^2} F_{\m \n}F^{\m \n} +|D_\m \phi^a|^2 +m^2 |\phi^a|^2+\frac{\l}{2N_f} (\phi^a {\phi^a}^*)^2 
\end{align} 
with the covariant derivative $D_\m =\pa_u+iA_\m$. This theory has an interesting phase structure with a broken and unbroken phase which are separated by a second order phase transition. Both these Higgs and Coulomb phases are gapless. In the Coulomb (unbroken) phase there is a massless photon and massive charged scalars $\phi^a$, whereas in the Higgs (broken) phase there are $2N_f-2$ massless Goldstone bosons and the photon becomes massive. 

\begin{figure}[t]
\centering
\includegraphics[clip,height=4.0cm]{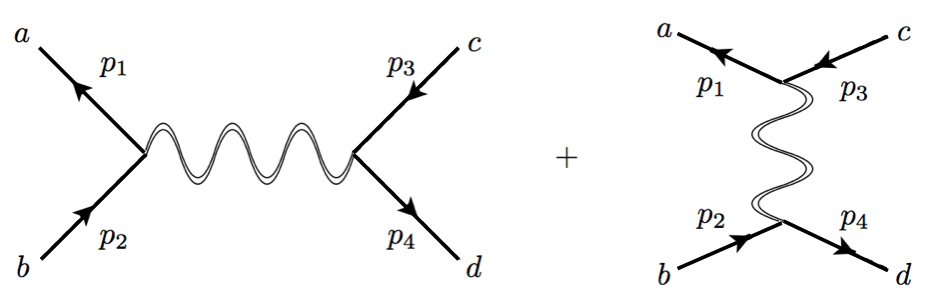}
\caption{The two diagrams that contribute to the $2\to 2$ scattering amplitude $\phi^a \phi^b \to {\phi^c}^* \phi^d$ at order $\frac{1}{N_f}$. The letters $a,b,c,d$ denote $SU(N_f)$ flavor indices.} \label{fig:figd33478} 
\end{figure}

At large-$N_f$ it is useful to introduce a Hubbard-Stratonovich field $\s$, and the lagrangian Eq.~\ref{eq:vkdfk4} becomes:
 \begin{align} 
L= \frac{1}{4e^2} F_{\m \n}F^{\m \n} +|D_\m \phi^a|^2 +m^2 |\phi^a|^2+\frac{\s}{\sqrt{N_f}} \phi^a {\phi^a}^*  -\frac{\s^2}{2\l}
\end{align} 
The equation of motion for $\s$ is $\s=\frac{\l}{\sqrt{N}_f}$. At order $1/N_f$ the propagators of $\s$ and the photon $A_\m$ can be computed exactly in the coupling constants $\l$ and $e$. The exact photon propagator is given by a resummation of the 1PI 2-point bubble diagrams, Fig.~\ref{fig:figd33438}. In momentum space, this resummation of bubble diagrams just a geometric series\footnote{In the Higgs phase, we have a non-zero vev $\phi^a\phi^a{}^*=N_f\Phi^2$,  the photon has a mass $m_A=\sqrt{2}e \Phi$, and there are massless Goldstones $M^2=0$. Instead of Eq.~\ref{eq:f8df}, the exact propagator in the Higgs phase is:\begin{align} \label{eq:dfkh43}
\langle A_\m(p) A_\n(-p)  \rangle = \frac{1}{N_f} \Big[ \frac{\a}{p^2+m_A^2+\alpha B^{(1)}(p^2,0)} \Big( \d_{\m \n} -\frac{p_\m p_\n}{p^2}\Big)+\zeta \frac{p_\m p_\n}{p^4}\Big]
\end{align} At the fixed point CFT, we have $M^2=0$ and $|\Phi|^2=0$, and the exact photon propagator is Eq.~\ref{eq:dfkh43} but with $m_A=0$. For more details, see e.g \cite{Ankur:2023lum}.}:
\begin{align} 
\label{eq:f8df}
\langle A_\m(p) A_\n(-p)  \rangle = \frac{1}{N_f} \Big[ \frac{\a}{p^2+\alpha B^{(1)}(p^2,M^2)} \Big( \d_{\m \n} -\frac{p_\m p_\n}{p^2}\Big)+\zeta \frac{p_\m p_\n}{p^4}\Big]
\end{align}
where $\a \equiv N_f e^2$ and $\zeta= N_f \xi$.  Where  $B^{(1)}(p^2, M^2)$ is the 1-loop bubble function, which  enters the 2-point function of currents:
\begin{align} 
\label{eq:kjhdf7}
\langle j_\m(p) j_\n(-p)  \rangle = -N_f B^{(1)}(p^2,M^2) \Big( \d_{\m \n} -\frac{p_\m p_\n}{p^2}\Big)
\end{align} 
$B^{(1)}(p^2,M^2)$ is a simple function, easily computable as a 1-loop Feynman diagram. The $2\to 2$ scattering amplitude ${\phi^a}^* \phi^b \to {\phi^c}^* \phi^d$, will simply be the exchange diagram with exchange of the exact photon propagator Eq.~\ref{eq:f8df} and Fig~\ref{fig:figd33478}:
\begin{align} 
&\mm{M}_{ab\to cd} = \frac{1}{N_f} \Big( \d^{ab}\d^{cd}\mm{M}(s,t) + \d^{ac}\d^{bd}\mm{M}(t,s)\Big)
\nn
&\mm{M}(s,t) =   \frac{\a(s-4M^2+2t)}{s-\alpha B^{(1)}(-s,M^2)} \Big( \d_{\m \n} -\frac{p_\m p_\n}{p^2}\Big)+\zeta \frac{p_\m p_\n}{p^4} 
\end{align} 
The poles of the scattering amplitude $\mm{M}(s,t)$ are thus determined by the zeros of the denominator, i.e $s-\alpha B^{(1)}(-s,M^2)=0$.

\subsection{Large-$N_f$ scalar QED on Anti de-Sitter space}

 We now consider the same model, but defined on $AdS_{d+1}$ instead of Euclidean space. We impose a Dirichlet boundary condition at the boundary of AdS, which gives rise to a global symmetry on the boundary. Just as in flat space, we would want to find the exact photon propagator at order $1/N_f$. In flat-space, working in momentum space enabled to resum bubble diagrams in a geometric series, Eq.~\ref{eq:f8df}. In AdS, the spectral representation can be viewed as the analogue of momentum space, and the spectral representation will enable us to resum bubble diagrams, just as in Fig.~\ref{fig:figd33438}. For more details, see e.g \cite{Ankur:2023lum}. The two-point function of the $U(1)$ conserved currents (the analogue of Eq.~\ref{eq:kjhdf7}):
 \begin{align} 
\langle j_M(X) j_N(Y)  \rangle = - \int_{-\infty}^\infty d\n B^{(1)}(\n) \O^{(1)}_{\n M N}(X,Y)
\end{align} 
where $B^{(1)}(\n)$ is the 1-loop bubble in the spectral representation, and $\O^{(1)}_{\n M N}(X,Y)$ is the spin-1 AdS harmonic function. The sum of bubble diagrams becomes a geometric series in spectral space $\n$, due to how $\O^{(1)}_{\n M N}(X,Y)$ behaves under convolution. Using this fact, the exact photon propagator is (the analogue of Eq.~\ref{eq:f8df}), Fig.~\ref{fig:figd33438}:
\begin{align} 
\label{eq:fjhdfv8}
\langle A_M(X) A_N(Y)  \rangle = \frac{1}{N_f}  \int d\n \frac{1}{\n^2+(\frac{d}{2}-1)^2+\a B^{(1)}(\n)} \O^{(1)}_{\n M N}(X,Y) + \nabla^M_X \nabla^N_y L(u)
\nn
\end{align} 
The 4-point boundary correlator is then obtained by attaching external legs, giving exchange Witten diagrams, with exchange of the exact propagator Eq.~\ref{eq:fjhdfv8}:
\begin{align} 
\langle \phi^a(P_1)  \phi^b{}^*(P_2)\phi^c{}^*(P_3)\phi^d(P_4) \rangle |_{O(\frac{1}{N_f})} = \frac{1}{N_f} \Big( \d^{ab}\d^{cd} g_{12|34} + \d^{ac}\d^{bd} g_{13|24} \Big)
\end{align} 
where
\begin{align} 
g_{12|34}=
  \int d^{d+1}X d^{d+1}Y K_{\D}(P_1,X) i \nabla_M^X  K_{\D}(P_2,X)  K_{\D}(P_3,Y) i \nabla_N^Y  K_{\D}(P_4,Y)  \langle A_M(X) A_N(Y)  \rangle
  \nn
\end{align} 
Where $K_{\D}$ are bulk-to-boundary propagators.
Using the split and spectral representation Eq.~\ref{eq:fjhdfv8}, the 4-point function becomes\footnote{where the factor $A$ is given below Eq.~\ref{eq:98d3}.} $g_{12|34}= A\times g_{l=1}(z, \bar z)$:
\begin{align} 
\label{eq:dfjk34}
g_{l=1}(z, \bar z) = 
\int_{-\infty}^\infty d\n  \frac{1}{\n^2+(\frac{d}{2}-1)^2+\a B^{(1)}(\n)}  \Upsilon_{\nu,l}^{\D}   \mm{K}^\D_{\frac{d}{2}+i\n,1} (z,\bar z) 
\end{align} 
Now we will see an important case in which we can compute this integral and obtain the exact 4-point function $g_{l=1}(z, \bar z)$ in terms of a different tree-level 4-point function.

\subsubsection{Computing the 4-point correlator at the bulk conformal point}
\begin{figure}[t]
\centering
\includegraphics[clip,height=6.7cm]{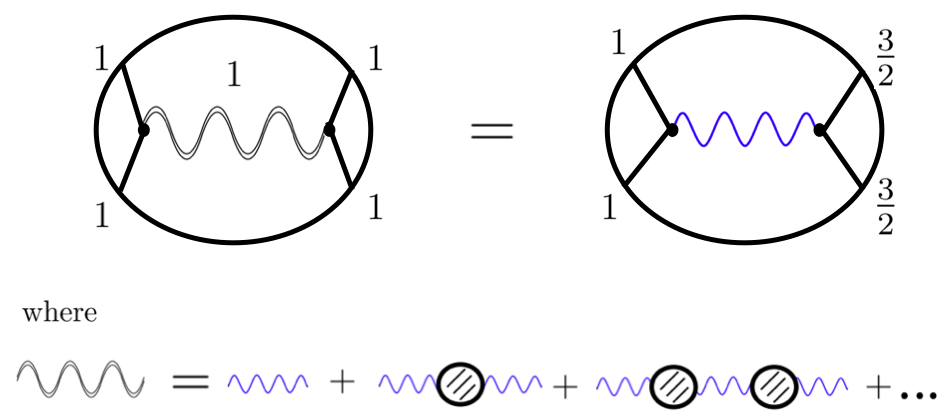}
\caption{Showing Eq.~\ref{eq:fnbvf8}. For scalar QED at the bulk conformal point (i.e $\D=1$ and $\a\to \infty$), the exact resummed 4-point function is equal to the tree-level photon exchange diagram on the right.} \label{fig:figd334} 
\end{figure}

Focusing on $AdS_3$ ($d=2$), the 1-loop bubble is \cite{Ankur:2023lum}:
\begin{align} 
\label{eq:vndkkf5}
B^{(1)}(\n) = \frac{\n}{16\pi (\n^2+1)} \Big( -2(2\D-3)\n +(\n^2 +4(\D-1)^2)(i\psi(\D-\frac{i \n}{2})- i\psi(\D+\frac{i \n}{2})) \Big)
\nn
\end{align} 
where $\psi(x)$ is the digamma function. In \cite{Ankur:2023lum}, we showed strong proof that a bulk conformal point exists when $\D=1$ and $\a \to \infty$. Plugging $\D=1$, the 1-loop bubble function simplifies:
\begin{align} 
\label{eq:dfnckd8}
B^{(1)}(\n) = \frac{\n^3}{16(\n^2+1)} \coth (\frac{\pi \n}{2}) =  -\frac{i \n^3}{16(\n^2+1)} \frac{\G_{1+\frac{i \n}{2} } \G_{-\frac{i \n}{2} } }{\G_{\frac{1}{2}- \frac{i \n}{2} } \G_{\frac{1}{2}+ \frac{i \n}{2} }  }
\end{align}  
Thus, Eq.~\ref{eq:dfjk34} becomes at the bulk conformal point ($\a \to \infty$ and $\D \to 1$):
\begin{align} 
\label{eq:dfnckd82}
g_{l=1}(z, \bar z) = \frac{1}{\a} \int d\n   \frac{1}{B^{(1)}(\n))} \Upsilon_{\nu,l}^{\D}   \mm{K}^\D_{\frac{d}{2}+i\n,1} (z,\bar z) 
\end{align} 
Using simple gamma function identities, we get that:
\begin{align} 
\G_{ 1-\frac{i\n }{2}} \G_{1+\frac{i\n }{2}}  (B^{(1)}(\n))^{-1} =\frac{32}{\n^2}  \times \G_{ \d+ \frac{-\frac{d}{2}-i\n+1 }{2}}  \G_{ \d+\frac{-\frac{d}{2}+i\n+1 }{2}}\Big|_{\d=\frac{3}{2}, d=2} 
\end{align} 
and hence Eq.~\ref{eq:dfnckd82} becomes
\begin{align} 
\label{eq:fnbvf8}
g_{l=1}(z, \bar z) = 
\frac{32}{\a} \int d\n  \frac{1}{\n^2 }  \Upsilon_{\nu,l=1}^{a_p=1,b_p=\frac{3}{2}}   \mm{K}^\D_{\frac{d}{2}+i\n,1} (z,\bar z) 
\end{align} 
The RHS in Eq.~\ref{eq:fnbvf8} can be seen to be the spectral representation of a tree-level exchange diagram of a photon! The $\frac{1}{\n}$ is the propagator of a photon, and the external scaling dimensions of the tree-level diagram are $\D_1=\D_2=1$ and $\D_3=\D_4=\frac{3}{2}$, see Fig.~\ref{fig:figd334}. We have thus managed to compute the non-perturbative 4-point function in terms of a tree-level exchange diagram. In \cite{Carmi:2019ocp,Carmi:2021dsn}, for the $O(N)$ and Gross-Neveu model at large-$N$ at the bulk conformal point, we showed that the non-perturbative 4-point function is equal to a tree-level contact diagram. The result of Eq.~\ref{eq:fnbvf8} is analogous to those results, now for the case of large-$N_f$ scalar QED on AdS.

\section{2-point bulk correlators} 
\label{sec:se4} 
 
 In this work we are considering mainly 4-point boundary correlator functions (with the external legs ending on the boundary of AdS). The boundary theory has conformal symmetry, and therefore the space-time dependence of the 2-point and 3-point boundary correlators are completely fixed by the conformal symmetry. In this section we consider bulk 2-point correlators, e.g $\langle \phi(x_1) \phi(x_2) \rangle_{bulk}$, where the two external points $x_{1,2}$ end in the bulk (see the RHS of Fig.~\ref{fig:figd345} ). The bulk correlators are not constrained by the full conformal symmetry, and they have non-trivial dependence on the chordal distance between the two points $x_{1,2}$. We wish to deduce this dependence in this section.
  
 \subsection{Bulk scalar 2-point correlators}
 
 The bulk-to-bulk propagator of a scalar field with scaling dimension $\D$ is (Eq.~\ref{eq:fkjdf3}):
\begin{align} 
G_{\D}(x_1,x_2) =   \frac{\G_\D}{2\pi^{\frac{d}{2}} \G_{\D-\frac{d}{2}+1}\sqrt{\zeta(\zeta+4)}} \zeta^{-\D} {}_2F_1 (\D,\D-\frac{d-1}{2},2\D-d+1,-4 \zeta^{-1})
\nn
\end{align} 
Where  $\zeta$ is the chordal distance squared. It will be convenient to express the hypergeometric function above in terms of a Legendre-Q function. Then the propagator can be written as:
\begin{align} 
G_{\D}(x_1,x_2) = 
e^{-\pi i\frac{d-1}{2}} \pi^{-\frac{d+1}{2}}   ( \zeta (\zeta+4))^{-\frac{d+1}{2}} Q^{\frac{d-1}{2}}_{\D-\frac{d+1}{2}} ( 1+\frac{\zeta}{2})
\end{align} 
where $Q_\mu^\nu(x)$ is the associated Legendre function of the second kind. Thus the spectral representation of a general bulk 2-point function is (see Eq.~\ref{eq:nseeeeb}):
\begin{align} 
&g^{(2)}(\z) = \int_{-\infty}^\infty d\n F_\n \O_\n (x_1,x_2) =   \int_{-\infty}^\infty d\n F_\n i \n G_{\frac{d}{2}+i\n }
 \nn
&=\frac{1}{2\pi^{\frac{d}{2}} }  \frac{\zeta^{-(\frac{d}{2} )}}{\sqrt{\zeta(\zeta+4)}}  \int_{-\infty}^\infty d\n F_\n   \zeta^{-i \n} {}_2F_1 (\frac{d}{2}+i\n, i\n+\frac{1}{2},2i\n+1,-4 \zeta^{-1}) 
\end{align} 
Or in terms of $Q$:
\begin{align} 
\label{eq:polf4}
g^{(2)}(\z)  = e^{-\pi i\frac{d-1}{2}} \pi^{-\frac{d+1}{2}}   ( \zeta (\zeta+4))^{-\frac{d+1}{2}} \int_{-\infty}^\infty d\n F_\n  i\n    Q^{\frac{d-1}{2}}_{i\n-\frac{1}{2}} (1+\frac{\zeta}{2}) 
\end{align} 
 
\subsection{A relation to the 4-point correlator}

\begin{figure}[t]
\centering
\includegraphics[clip,height=4.2cm]{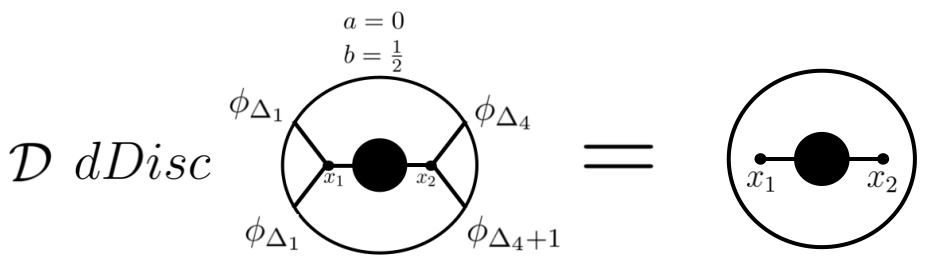}
\caption{ Relating the bulk two-point correlator on the right, to the $dDisc$ of the four-point function on the left. The black blob is the same on both sides, and is a general bulk two-point function. } \label{fig:figd345} 
\end{figure}

We will show in this subsection that the bulk 2-point correlator \ref{eq:polf4} is closely related to a specific boundary 4-point correlator.
 For $a=0$ and $b=\frac{1}{2}$, the scalar $J=0$ conformal block is given, for any $d$, by a hypergeometric function \cite{Dolan:2000ut}:
\begin{align} 
 \mathcal{K}^{\D_i}_{\frac{d}{2}+i\nu} (z,\bar z) = \frac{1}{\sqrt{v}} \Big(\frac{1+\sqrt{v}}{2} \Big)^{1-\D} {}_2 F_1 \Big(\frac{\D-1}{2},\frac{\D}{2},\D+1-\frac{d}{2}: \frac{u}{(1+\sqrt{v})^2} \Big)
\end{align} 
where the cross ratios are $u= z \bar z$ and $v=(1-z)(1-\bar z)$.
We can also write this in terms of the Legendre function:
\begin{align} 
&\mathcal{K}^{\D_i}_{\frac{d}{2}+i\nu} (z,\bar z) =     \b \times 4^\D \frac{\G_{\D-\frac{d}{2}+1}}{\G_{\D-1}} Q^{\frac{d-3}{2}}_{\D-\frac{d+1}{2}} (\sqrt{\frac{4}{X}})
\end{align} 
where $X\equiv \frac{4u}{(1+\sqrt{v})^2}$  and $\b \equiv  \Big(2^{-\frac{d-1}{2}} e^{-\frac{d-3}{2}\pi i} (1-\frac{X}{4})^{-\frac{d-3}{4}} (\frac{X}{4})^{\frac{d-1}{4}} \Big)$.
Additionally, the $\Upsilon$ factor of Eq.~\ref{eq:sdskkf} for $a=0$, $b=\frac{1}{2}$ is:
\begin{align} 
\Upsilon_\nu^{\D_i} = \pi \Big( \frac{\G_{a_p -\frac{d}{2} +\frac{\D}{2}} \G_{b_p -\frac{d}{2} +\frac{\D}{2}}}{\G_{1-a_p  +\frac{\D}{2}} \G_{1-b_p  +\frac{\D}{2}}}\Big)  2^{3-2\D} \frac{\G_{\D-1}}{\G_{\D-\frac{d}{2}}}
\end{align} 
 with $\D=\frac{d}{2}+i\n$.
Therefore
\begin{align} 
\Upsilon_\nu^{\D_i}  \mathcal{K}^{\D_i}_{\frac{d}{2}+i\nu} (z,\bar z)   = 8\pi \b  \Big( \frac{\G_{a_p -\frac{d}{2} +\frac{\D}{2}} \G_{b_p -\frac{d}{2} +\frac{\D}{2}}}{\G_{1-a_p  +\frac{\D}{2}} \G_{1-b_p  +\frac{\D}{2}}}\Big)   (\D-\frac{d}{2}) Q^{\frac{d-3}{2}}_{\D-\frac{d+1}{2}} (\sqrt{\frac{4}{X}})
\end{align} 
Now we can make the factor in the brackets be 1, if we choose either $a_p=b_p =\frac{d+2}{4}$ or $b_p= 1-a_p+\frac{d}{2}$. So we get:
\begin{align} 
\Upsilon_\nu^{\D_i}  \mathcal{K}^{\D_i}_{\frac{d}{2}+i\nu} (z,\bar z)   = 8\pi \b   (i\n) Q^{\frac{d-3}{2}}_{i\n-\frac{ 1}{2}} (\sqrt{\frac{4}{X}})
\end{align} 
Plugging this in Eq.~\ref{eq:fdf33} gives:
\begin{align} 
 dDisc[g_{l=0}] = \int_{-\infty}^\infty d\n F_\n  \Upsilon_\nu^{\D_i}  \mathcal{K}^{\D_i}_{\frac{d}{2}+i\nu} (z,\bar z) = 8\pi \b   \int_{-\infty}^\infty d\n  (i\n) Q^{\frac{d-3}{2}}_{i\n-\frac{ 1}{2}} (\sqrt{\frac{4}{X}}) 
\end{align} 
 Now we can use an identity for the Legendre function:
\begin{align} 
  \mm{D}   Q^{\frac{d-3}{2}}_{i\n-\frac{1}{2}} (x) =  Q^{\frac{d-1}{2}}_{i\n-\frac{1}{2}} (x)
\end{align} 
where the differential operator is defined as:
\begin{align} 
 \mm{D} \equiv - (1-x^2)^{\frac{d+1}{4}}  \frac{d}{dx}  (1-x^2)^{\frac{d-1}{4}}
\end{align} 
 Therefore
\begin{align} 
\label{eq:cvd4}
 \mm{D}[ dDisc[g_{l=0}] ]= \int_{-\infty}^\infty d\n  F_\n \Upsilon_\nu^{\D_i}   \mm{D}  \mathcal{K}^{\D_i}_{\frac{d}{2}+i\nu} (z,\bar z) = 8\pi \b   \int_{-\infty}^\infty d\n  (i\n) F_\n Q^{\frac{d-1}{2}}_{i\n-\frac{ 1}{2}} (\sqrt{\frac{4}{X}}) 
\end{align} 
Comparing Eqs.~\ref{eq:cvd4} and \ref{eq:polf4}, we get the relation:
\begin{align} 
\boxed{  \mm{D} dDisc[   g_{l=0}] ]= c_2 g^{(2)}(\z) \big|_{1+\frac{\zeta}{2} \to \sqrt{\frac{4}{X}}) } }
\end{align} 
where $c_2 \equiv \frac{8\pi \b}{e^{-\pi i\frac{d-1}{2}} \pi^{-\frac{d+1}{2}}   ( \zeta (\zeta+4))^{-\frac{d+1}{2}}}$. In other words, these 2-point and 4-point functions above have the same spectral integrals, and thus are directly related to each other. This identity is shown in Fig.~\ref{fig:figd345}.

\subsection{Bulk spin-1 2-point correlators}

\subsubsection{spin-1 propagators}

Consider a bulk spin-1 massive Proca field with mass $m^2 = (\Delta-1)(\Delta-d+1)$. The propagator $G_{\Delta,1}$ has two structures \cite{Costa:2014kfa}:
\begin{equation}
\label{W5}
G_{\Delta,1} \left( x_1, x_2 ; W_1, W_2\right)=\left(W_1 \cdot W_2\right) (G_{\Delta,1})_0\left(u\right)  +\left(W_1 \cdot x_2\right)\left(W_2 \cdot x_1\right) (G_{\Delta,1})_1\left(u\right) ~,
\end{equation}
where the coefficient functions $(G_{\Delta,1})_0(u)$ and $(G_{\Delta,1})_1(u)$ are
\begin{align}
\label{eq:fjks7d}
\begin{split}
(G_{\Delta,1})_0\left(u\right) & =(d-\Delta) f_1(u)-\frac{1+u}{u} f_2(u), \\
(G_{\Delta,1})_1\left(u\right) & = \frac{(1+u)(d-\Delta)}{u(2+u)} f_1(u)-\frac{d+(1+u)^2}{u^2(2+u)} f_2(u)~.
\end{split}
\end{align}
and where we defined the two hypergeometric functions:
\begin{align}
\begin{split}
 f_1(u) & =\mathcal{N}(2 u)^{-\Delta}{ }_2 F_1\left(\Delta, \frac{1-d+2 \Delta}{2}, 1-d+2 \Delta,-\frac{2}{u}\right)~, \\
 f_2(u) & =\mathcal{N}(2 u)^{-\Delta}{ }_2 F_1\left(\Delta+1, \frac{1-d+2 \Delta}{2}, 1-d+2 \Delta,-\frac{2}{u}\right)~, \\
& \mathcal{N}  \equiv \frac{\Gamma(\Delta+1)}{2 \pi^{d / 2}(d-1-\Delta)(\Delta-1) \Gamma\left(\Delta+1-\frac{d}{2}\right)}~.
\end{split}
\end{align}
See also \cite{Ankur:2023lum}. In terms of the Legendre function, the coefficients can be written as follows:
\begin{align} 
&(d-\D)f_1(u)= 
2^{\frac{-1-d}{2}}e^{-\pi i\frac{d-1}{2}} \pi^{-\frac{d+1}{2}}   (u (u+2))^{-\frac{d-1}{4}} \frac{\n^2+\frac{d^2}{4}}{\n^2+(\frac{d}{2}-1)^2} Q^{\frac{d-1}{2}}_{i\n-\frac{1}{2}} (1+u) 
\nn
&f_2(u)= -  2^{\frac{-1-d}{2}}e^{-\pi i\frac{d+1}{2}} \pi^{-\frac{d+1}{2}}   u(u (u+2))^{-\frac{d+1}{4}}\frac{1}{\n^2+(\frac{d}{2}-1)^2} Q^{\frac{d+1}{2}}_{i\n-\frac{1}{2}} (1+u) 
\end{align} 
where $u=\frac{\zeta}{2}$ and $\D= i\n +\frac{d}{2}$.

\subsubsection{2-point bulk correlators}

The propagator is the tree-level bulk 2-point correlation function. Here we want to consider a more general bulk correlator. We will show that spin-1 bulk 2-point functions can be computed from spin-0 bulk 2-point functions.

If we have a spin-$1$ field in $AdS_{d+1}$. A general bulk 2-point function has the following spectral representation:
 \begin{align} 
&g_2(x_1,x_2,W_1,W_2) =
 (W_1 \cdot \nabla_1) (W_2 \cdot \nabla_2)) \int_{-\infty}^\infty d\n \tilde{F}^L(\n) \O_{\n,0}(x_1,x_2:W_1,W_2)
 \nn
&+ \int_{-\infty}^\infty d\n \tilde{F}^\perp(\n) \O_{\n,1}(x_1,x_2:W_1,W_2)
\end{align} 
where $\O_{\n,0}$  and $\O_{\n,1}$ are the spin-0 and spin-1 AdS harmonic functions respectively, see \cite{Costa:2014kfa,Ankur:2023lum}. We focus on the transverse part in the second line. Now the AdS harmonic function is:
\begin{align} 
\O_{\n,1} (x_1,x_2,W_1,W_2) = \frac{i \n}{2 \pi } \Big(  G_{\frac{d}{2}+i\n,1} (x_1,x_2,W_1,W_2) - G_{\frac{d}{2}-i\n,1} (x_1,x_2,W_1,W_2) \Big)
\nn
\end{align} 
Plugging this in the transverse part gives:
\begin{align} 
 &\int_{-\infty}^\infty d\n \tilde{F}^\perp(\n) \O_{\n,1}=  \frac{1}{2\pi} \int_{-\infty}^\infty d\n \tilde{F}^\perp(\n)  i\n  G_{\frac{d}{2}+i\n,1} = 
 \nn
& \left(W_1 \cdot W_2\right) \int_{-\infty}^\infty d\n \tilde{F}^\perp(\n)  i\n (G_{\frac{d}{2}+i\n,1})_0\left(u\right)  +\left(W_1 \cdot x_2\right)\left(W_2 \cdot x_1\right)  \int_{-\infty}^\infty d\n \tilde{F}^\perp(\n)  i\n (G_{\frac{d}{2}+i\n,1})_1\left(u\right)
\end{align} 
From Eq.~\ref{eq:fjks7d}, this gives rise to two terms:
\begin{align} 
\label{eq:gh2323}
 & \int d\n \tilde{F}(\n) i \n  (-i\n+\frac{d}{2}) f_1(u) 
 \nn
& = 2^{\frac{-1-d}{2}}e^{-\pi i\frac{d-1}{2}} \pi^{-\frac{d+1}{2}}   (u (u+2))^{-\frac{d-1}{4}} \int d\n F_\n i \n  \frac{\n^2+\frac{d^2}{4}}{\n^2+(\frac{d}{2}-1)^2} Q^{\frac{d-1}{2}}_{i\n-\frac{1}{2}} (1+u)
\nn
& = 2^{\frac{-1-d}{2}}e^{-\pi i\frac{d-1}{2}} \pi^{-\frac{d+1}{2}}   (u (u+2))^{-\frac{d-1}{4}} \int d\n F_\n i \n  \Big[ 1+\frac{d-1}{\n^2+(\frac{d}{2}-1)^2}\Big] Q^{\frac{d-1}{2}}_{i\n-\frac{1}{2}} (1+u)
\end{align} 
and
\begin{align} 
\label{eq:gh232}
& \int d\n F_\n i \n   f_2(u) 
 \nn
& =  - 2^{\frac{-1-d}{2}}e^{-\pi i\frac{d+1}{2}} \pi^{-\frac{d+1}{2}}   u(u (u+2))^{-\frac{d+1}{4}}\int d\n F_\n i \n   \frac{1}{\n^2+(\frac{d}{2}-1)^2} Q^{\frac{d+1}{2}}_{i\n-\frac{1}{2}} (1+u)
\end{align} 
Defining the differential operator $\widehat{\mm{D}}  \equiv - (1-x^2)^{\frac{d+3}{4}}  \frac{d}{dx}  (1-x^2)^{\frac{d+1}{4}}$, that acts on the Legendre function as follows:
 \begin{align} 
\widehat{\mm{D}}   Q^{\frac{d-1}{2}}_{i\n-\frac{1}{2}} (x) =  Q^{\frac{d+1}{2}}_{i\n-\frac{1}{2}} (x)
\end{align} 
therefore we get:
\begin{align} 
\label{eq:cndso4}
& \int d\n F_\n i \n   f_2(u) 
 \nn
& = - 2^{\frac{-1-d}{2}}e^{-\pi i\frac{d+1}{2}} \pi^{-\frac{d+1}{2}}   u(u (u+2))^{-\frac{d+1}{4}}  \widehat{\mm{D}}  \int d\n F_\n i \n   \frac{1}{\n^2+(\frac{d}{2}-1)^2} Q^{\frac{d-1}{2}}_{i\n-\frac{1}{2}} (1+u)
\end{align}
These two integrals can be easily related to scalar bulk 2-point integrals given in Eq.~\ref{eq:polf4}. In particular to get the integral in Eq.~\ref{eq:cndso4}, one simply attaches a free propagator. To get the integral in Eq.~\ref{eq:gh2323}, one attaches a free propagator and acts with the derivative operator $2D_{z,\bar z}$ with $l=0$, see Eq.~\ref{eq:dfdd}. Therefore the spin-1 bulk 2-point functions can be computed from spin-0 bulk 2-point functions.

\section{The 4-point 1-loop bubble diagram}
\label{sec:se6}

In this subsection we compute various 4-point 1-loop bubble diagrams in AdS. What we are after is the dependence of the 4-point function on the cross-ratios $z$ and $\bar z$. Therefore we will drop overall factors which do not depend on the cross-ratios.
Let us look at the 1-loop scalar bubble in the spectral representation. The 1-loop bubble $\tilde B$ has single poles, see \cite{Carmi:2018qzm}:
\begin{align} 
\label{eq:dhbdf9f}
\tilde{B}(\n)  \overset{\frac{d}{2}+i\n \sim 2\D+2n}{\sim}  \widehat{A}_{\D}
\L_{n,\D} \times \frac{1}{\frac{d}{2}+i\n -(2\D+2n)} 
\end{align}
where we defined the product of gamma functions: 
\begin{align} 
\L_{n,\D} \equiv \frac{\G_{\frac{d}{2}+n} \G_{n+\D}\G_{\frac{1}{2}-\frac{d}{2}+n+\D} \G_{2\D+n-\frac{d}{2}}}{\G_{n+1} \G_{n+\D+1-\frac{d}{2}} \G_{n+\D+\frac{1}{2}}\G_{1-d+n+2\D}}  
\end{align} 
 Therefore from Eq.~\ref{eq:df54}
\begin{align} 
\label{eq:jf9d}
dDisc [g_4(z, \bar z)] \Big|_{1-loop}=  \widehat{A}_{\D,l} \sum^\infty_{n=0} \L_{n,\D}\times  \Upsilon_{\nu,l=0}^{\D_i} \mathcal{K}^{\D_i}_{\frac{d}{2}+i\nu,l=0} (z,\bar z) \Big|_{i\n+\frac{d}{2} \to 2\D+2n}
\end{align} 
we will use this relation below.

 \subsection{Scalar bubble diagram, odd $d$}

\noindent $\bullet$ $d=3$

\begin{figure}[t]
\centering
\includegraphics[clip,height=3.3cm]{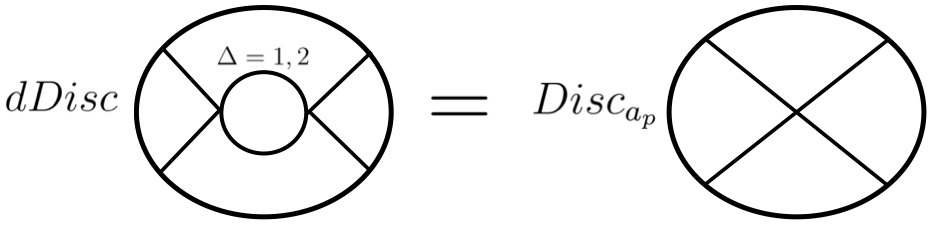}
\caption{Showing the relation of Eq.~\ref{eq:hjf342} between the 4-point 1-loop bubble diagram and contact diagrams for $d=3$ and $\D=1,2$. Similar relations exist in $d=5$ with $\D=3,4$, shown in Eq.~\ref{eq:kd999d1}-\ref{eq:kd999d2}.}\label{fig:figd55} 
\end{figure}

\noindent The poles of the 1-loop bubble in $d=3$ and $l=0$ are, see Eq.~\ref{eq:dhbdf9f}:
\begin{align} 
2 \tilde{B}_\n \sim -\frac{1}{\frac{d}{2} +i\n -(2\D+2n)} \sqrt{\pi}2^{1-4\D}\G_{2\D}\G_{2\D+1} \frac{\G_{n+\frac{3}{2}} \G^2_{\D+n+\frac{1}{2}}\G_{2\D+n+\frac{1}{2}}  }{\G_{n+1}  \G^2_{\D+n+1} \G_{n+2\D}}
\end{align} 
Thus for $\D=1,2$, the residue is constant:
\begin{align} 
2 \tilde{B}_\n \sim -\frac{\frac{4}{\sqrt{\pi}}}{\frac{d}{2} +i\n -(2\D+2n)} 
\end{align} 
Using Eq.~\ref{eq:jf9d}, \ref{eq:sdskkf} and \ref{eq:dfkss}, we see that for $b_p=\D$ and $\D=1,2$ we have
\begin{align} 
\label{eq:hjf342}
dDisc[g_{l=0}]\Big|_{1-loop}  = Disc_{a_p} [g_{l=0}(z, \bar z)] \Big|_{Contact} 
\end{align} 
Thus, this family of 1-loop bubble diagrams can be computed via the tree-level contact diagram. This is schematically shown in Fig.~\ref{fig:figd55}.\\

\noindent $\bullet$ $d=5$

\noindent  In $d=5$ and When $\D=3,4$, the residue is :
\begin{align} 
2 \tilde{B}_\n \sim -\frac{\frac{1}{3\sqrt{\pi}} (2n+3)(2n+4)}{\frac{d}{2} +i\n -(2\D+2n)}  \ \ \ \ \ \ , \ \ \ \ \ \ \ 2 \tilde{B}_\n \sim -\frac{\frac{1}{3\sqrt{\pi}} (2n+2)(2n+9)}{\frac{d}{2} +i\n -(2\D+2n)} 
\end{align} 
Using Eq.~\ref{eq:dfdd}, and after choosing $b_p=\D$, we get a relation between the 1-loop and contact diagram:
\begin{align} 
\label{eq:kd999d1}
dDisc[g_{l=0}] \Big|_{\D=3}= (2D_{z,\bar z}+6)  Disc_{a_p} [g_{l=0}(z, \bar z)] \Big|_{Contact} 
\end{align} 
and
\begin{align} 
\label{eq:kd999d2}
dDisc[g_{l=0}\Big|_{\D=4}= (2D_{z,\bar z}-6)  Disc_{a_p} [g_{l=0}(z, \bar z)] \Big|_{Contact} 
\end{align}

\subsection{Fermionic bubble diagram}

\begin{figure}[t]
\centering
\includegraphics[clip,height=3.8cm]{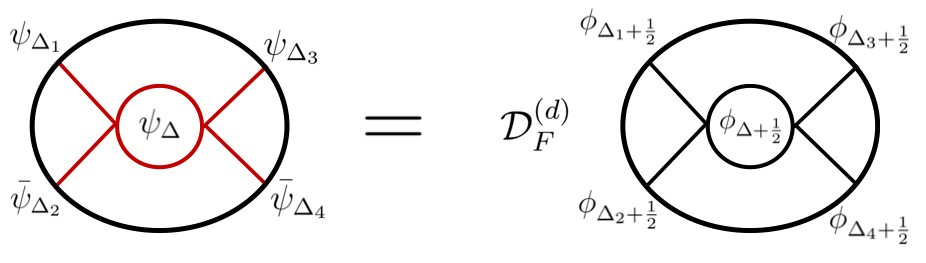}
\caption{Showing a relation between the 4-point 1-loop bubble diagrams of fermions and scalars in $d=1,2$. The relations are given in Eqs.~\ref{eq:pold3d1} and \ref{eq:pold3d2}.}\label{fig:figd545} 
\end{figure}

In section~6 of \cite{Carmi:2018qzm}, we studied the 1-loop fermionic bubble diagram in $d=1,2$. In the current subsection we relate the 4-point 1-loop fermionic bubble to the scalar bubble.\\

\noindent $\bullet$ $d=1$

\noindent  In $d=1$ the poles of the 1-loop fermion bubble are \cite{Carmi:2018qzm}:
\begin{align} 
2 \tilde{B}^{(F)}_\n \sim -\frac{1}{\frac{d}{2} +i\n -(2\D+2n+1)} \sqrt{\pi}2^{1-4\D}\G_{2\D}\G_{2\D+1} \frac{\G_{n+\frac{3}{2}} \G^2_{\D+n+\frac{1}{2}}\G_{2\D+n+\frac{1}{2}}  }{\G_{n+1}  \G^2_{\D+n+1} \G_{n+2\D}}
\end{align} 
The scalar bubble is given in Eq.~\ref{eq:dhbdf9f}. Thus, attaching bulk-to-boundary propagators, one can see that the 4-point 1-loop fermion bubble is directly related to the scalar 1-loop bubble with external scalars (see Fig.~\ref{fig:figd545}):
\begin{align} 
\label{eq:pold3d1}
 \boxed{
 g_F^{(\D_i,\D)} (z, \bar z)   = \mm{D}_F^{(d=1)}  g_B^{(\D_i+\frac{1}{2},\D+\frac{1}{2})} (z, \bar z)  }
\end{align} 
where we defined the differential operator $\mm{D}_F^{(d=1)} \equiv  c_\D( 2D_{z,\bar z}-2\D(2\D-1))$. \\

\noindent $\bullet$ $d=2$

\noindent  In $d=2$ the poles of the 1-loop fermion bubble are \cite{Carmi:2018qzm}:
\begin{align} 
2 \tilde{B}_\n^{(F)} \sim -\frac{1}{ \frac{d}{2}+  i\n -(2\D+2n+1)} \frac{(n+1)(2\D+n-1)}{n+\D}
\end{align} 
Thus, attaching bulk-to-boundary propagators, the 4-point 1-loop fermion bubble is directly related to the scalar 1-loop bubble with external scalars (see Fig.~\ref{fig:figd545}):
\begin{align} 
\label{eq:pold3d2}
 \boxed{
 g_F^{(\D_i,\D)} (z, \bar z)   = \mm{D}_F^{(d=2)}  g_B^{(\D_i+\frac{1}{2},\D+\frac{1}{2})} (z, \bar z)  }
\end{align} 
 where we defined the differential operator $\mm{D}_F^{(d=2)} \equiv   \tilde{c}_\D (2D_{z,\bar z}-4\D^2+8\D-3)$.

\section{4-point ladder diagrams}
\label{sec:se7}

 Ladder diagrams are an important class of diagrams in QFT, which occur in large-$N$ theories such as Chern-Simons Matter theories and 2-dimensional QCD (i.e the 't Hooft model). Ladder diagram are computationally more technically challenging than bubble diagrams (which can be reduced to the computation of the 1-loop bubble).
 
In this section we discuss the computation of ladder diagrams with spinning fields, 
in particular for gauge theories. This extends the results of \cite{Carmi:2021dsn}, in which we derived the spectral representation for ladder diagrams of scalar fields in AdS. We recursively obtain expressions for an $k$-loop ladder diagram, by "gluing" together the rungs of the ladder. Each such gluing results in a factor of a 6j symbol, and the $k$-loop ladder is given by a convolution of $(k+1)$ 6J-symbol factors.

\begin{figure}[t]
\centering
\includegraphics[clip,height=4.8cm]{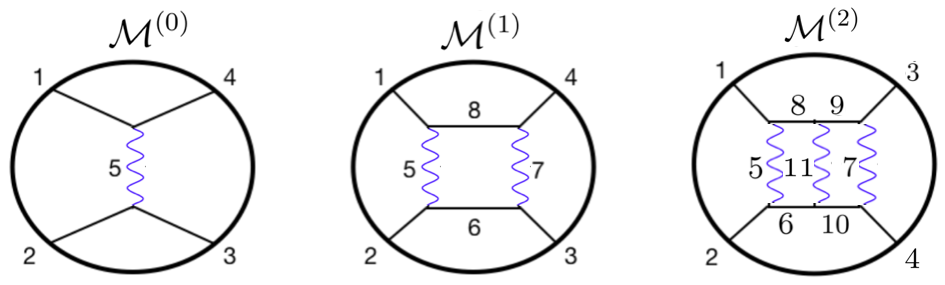}
\caption{Four-point ladder diagrams for exchanges of a spin-$L$ field. \textbf{Left:} The 0-loop ladder, tree level exchange. \textbf{Middle:} The 1-loop ladder, i.e the box diagram.  \textbf{Right:} The ladder diagram at 2-loops.}\label{fig:figd5} 
\end{figure}

We consider a theory with a cubic interaction of scalar field with a spin-$L$ field. The expression for the 0-loop ladder diagram, which is just the tree-level exchange diagram Fig.~\ref{fig:figd5}-Left is:
\begin{align}
\label{eq:lkd3}
\mm{M}^{(0)}=  \mm{M}^{3214}_{5 , exch} =\sum^\infty_{J =0} \int_{-\infty}^\infty \frac{d\m}{2\pi i}\ R^{1234}_{\mm{O}_5 ,\mm{O}_{\m,J}}  \ \Psi^{1234}_{ \m,J}
\end{align}
This is derived in Eq.~\ref{eq:lkd3}. Here $R^{1234}_{\mm{O}_5 ,\mm{O}_{\m,J}}$ is a sum of 6J symbols given in Eq.~\ref{eq:kdlsi}. 

We can now add a rung to the ladder, giving 1-loop the box diagram Fig~\ref{fig:figd5}-Middle. This diagram can then be computed by gluing two tree-level diagrams, schematically:
\begin{align}
\mm{M}^{(1)} = \mm{M}_{Box} =  \mm{M}^{621\tilde{\underline{8}}}_{5 , exch}  \otimes \mm{M}^{3\tilde{\underline{6}}\underline{8}4}_{7, exch} 
\end{align}
And the ladder diagram at $k$-loops is obtained by gluing $(k+1)$ tree-level diagrams:
\begin{align}
\mm{M}^{(k)} =  \mm{M}_{1,exch.} \otimes \mm{M}_{2,exch.} \otimes \cdots \otimes  \mm{M}_{k,exch.} \otimes \mm{M}_{k+1,exch.}
\end{align}
This procedure yields, for the $k$-loop ladder diagrams, the expression for the OPE function:
 \begin{align}
\mm{M}^{(k)}= \sum^\infty_{J_5 =0} \int_{-\infty}^\infty \frac{d\m_5}{2\pi i} C^{(k)}_{\m_5,J_5} \Psi^{1234}_{\m_5, J_5}
\end{align}
where $C^{(k)}_{\m_5,J_5}$is the respective OPE function at $k$-loops. From Eq.~\ref{eq:lkd3} we have for the tree-level exchange diagram:
\begin{align}
\label{eq:pldnnd666}
C^{(0)}_{\m,J}  =  R^{\zz{1},\zz{2},\zz{3},\zz{4}}_{ \m,J}    
\end{align}
The external points of the 4-point diagram are highlighted in red in our formulas. At 1-loop we have:
\begin{align}
\label{eq:flkkd4}
C^{(1)}_{\m,J}  =  B_{ \m,J } \int_{-\infty}^\infty  \bigg( \prod_{i=6,8} \frac{d \n_i \n_i^2   }{ \n_i^2+(\D_i-\frac{d}{2})^2 }  \bigg)  R^{\zz{1},\zz{2},6,8}_{ \m,J}   R^{\tilde 8,\tilde{6},\zz{3},\zz{4}}_{ \m,J}
\end{align}
We get the following expression for the $k$-loop OPE function of the ladder diagram:
\begin{align}
\label{eq:pol90}
\boxed{
C^{(k-loop)}_{\m,J} =\big( B_{ \m,J }\big)^k
 \int_{-\infty}^\infty  \bigg( \prod_{horiz.} \frac{d \n_i \n_i^2   }{ \n_i^2+(\D_i-\frac{d}{2})^2 }  \bigg)    \prod_{vertical}  R_{ \m,J}   }
\end{align}
Where we defined the product:
\begin{align}
\label{eq:plmd3}
 \prod_{vertical}   R_{ \m,J }= R^{\zz{1},\zz{2},A_1,B_1}_{ \m,J} R^{\tilde{B}_1, \tilde{A}_1,A_2,B_2}_{ \m,J} R^{\tilde{B}_2, \tilde{A}_2,A_3,B_3}_{ \m,J}\ \ \cdots \ R^{\tilde{B}_{\hat k-1}, \tilde{A}_{\hat k-1}, A_{\hat k},B_{\hat k}}_{\ \m,J}  R^{\tilde{B}_{\hat k}, \tilde{A}_{\hat k},\zz{3},\zz{4}}_{ \m,J}
\end{align}
In appendix~\ref{sec:A1} we derive Eq.~\ref{eq:pol90} explicitly for the 2-loop case. The same type of derivation extends to $k$-loops, giving Eq.~\ref{eq:pol90}.


\section{Discussion}
\label{sec:se8}

In this work we used the spectral representation as a tool to compute various families of loop diagrams in anti de-Sitter space, including blob diagrams and ladder diagrams. The spectral representation makes various identities explicit, enabling to related loop Witten diagrams to lower loop diagrams. We saw that it was often simpler to compute the double-discontinuity of a diagram. The conformal dispersion relation \cite{Carmi:2019cub}, and it's generalization to mixed correlators \cite{carmi3}, can then be used to reconstruct the full 4-point correlator.  

The spectral representation has recently been used for de-Sitter space, and it would be interesting to import our AdS methods to compute blob diagrams in deSitter. The $O(N)$ model at large-$N$  has recently been studied in de Sitter space \cite{DiPietro:2023inn}, can one extend these computations for the case of the fermionic Gross-Neveu model and to large-$N_f$ QED on deSitter. We also leave for future work the computation of loops of gluons and gravitons in AdS and dS.

In order to make progress on analytic loop computations of diagrams in AdS, it can be useful to consider some simplifying limits, such as the flat-space limit of AdS diagrams. Conformal blocks simplify in the large $d$ limit, which might enable to perform analytical computations using the spectral representation. Another simplifying limit is the diagonal limit $\bar z=z$, which is essentially a 1d limit.  \\

\textbf{Acknowledgements}: I thank Lorenzo Di Pietro and Ankur Ankur for discussions. This work was supported by the Israeli Science Foundation (ISF) grant number 1487/21, and by the MOST NSF/BSF physics grant number 2022726.


\appendix

\section{Ladder diagrams}
\label{sec:A1}

\subsection{Ladder diagram at tree-level}

The conformal partial wave (CPW) $\Psi$ in the s-channel is related to the CPW in the t-channel via, see \cite{Liu:2018jhs,Meltzer:2019nbs}:
\begin{align} 
\label{eq:fksppf}
\Psi^{3214}_{\m_6,l_6} (x_i)=\sum^\infty_{J =0} \int_{-\infty}^{\infty} \frac{d\m}{2\pi i}       \left\{ \begin{array}{c} \ \mm{O}_1,\mm{O}_2,\mm{O}_{\m_6,l_6} \ \  \\   \mm{O}_3,\mm{O}_4,\mm{O}_{\m,J } \end{array}\right\} \frac{1}{n_{\mm O_{\m,J}}} \Psi^{1234}_{\m,J}
\end{align} 
Where the factor in the brackets is a 6J symbol.
\begin{align} 
\left\{ \begin{array}{c} \ \mm{O}_1,\mm{O}_2,\mm{O}_{\m_6,l_6} \ \  \\   \mm{O}_3,\mm{O}_4,\mm{O}_{\m,J } \end{array}\right\}  = K^{14}_{\tilde 6} \left( \begin{array}{c} \ \mm{O}_1,\mm{O}_2,\mm{O}_{\m_6,l_6} \ \  \\   \mm{O}_3,\mm{O}_4,\mm{O}_{\m,J } \end{array}\right) +K^{23}_{ 6} \left( \begin{array}{c} \ \mm{O}_1,\mm{O}_2, \tilde{\mm{O}}_{\m_6,l_6} \ \  \\   \mm{O}_3,\mm{O}_4,\mm{O}_{\m,J } \end{array}\right)  
\end{align} 
Now consider the tree-level exchange diagram of a spin-$L$ field in AdS (Fig.~\ref{fig:figd5}-Left). Expanded in the direct $s$-channel partial waves, we have \cite{Costa:2014kfa}:
\begin{align}
\mm{M}^{3214}_{5 , exch} = A \sum_{l_5=0}^{L}  \int_{-\infty}^{\infty} d \m_5 b_{l_5}(\m_5) \Psi^{3214}_{\m_5,l_5}
\end{align}
where $A$ is defined in Eq.~\ref{eq:nvjdf8}, and $b_{l_5}(\m_5)$ are the OPE functions for the 4-point correlator. Using Eq.~\ref{eq:fksppf} gives an expansion in terms of the $t$-channel CPWs:
\begin{align}
\label{eq:fmv9f}
\mm{M}^{3214}_{5 , exch} =A  \sum^\infty_{J =0} \int \frac{d\m}{2\pi i} \sum_{l_5=0}^{L} \int d \m_5 b_{l_5}(\m_5)   K^{14}_{\tilde 5}    \left(\begin{array}{c} \mm{O}_1,\mm{O}_2,\mm{O}_{5,l_5}\ \  \\   \mm{O}_3,\mm{O}_4,\mm{O}_{\m,J } \end{array}\right) \frac{1}{n_{\mm O_{\m,J}}} \Psi^{1234}_{\m,J}
\end{align}
In the integrand above, we can define the following factor:
\begin{align}
\label{eq:kdlsi}
R^{1234}_{ \m,J} \equiv  \sum_{l_5=0}^{L} \int_{-\infty}^{\infty} d \m_5   b_{l_5}(\n) K^{14}_{\tilde 5}\left(\begin{array}{c} \mm{O}_1,\mm{O}_2,\mm{O}_{5,l_5}\ \ \  \\   \mm{O}_3,\mm{O}_4,\mm{O}_{\m ,J }  \end{array}\right)  \frac{1}{n_{\mm O_{\m,J}}}
\end{align}
Therefore Eq.~\ref{eq:fmv9f} gives:
\begin{align}
\label{eq:lkd3}
\mm{M}^{(0)}=  \mm{M}^{3214}_{5 , exch} =\sum^\infty_{J =0} \int_{-\infty}^{\infty}  \frac{d\m}{2\pi i}\ R^{1234}_{\m, J}  \ \Psi^{1234}_{ \m,J}
\end{align}
This is the expression for the tree-level 4-point ladder diagram, Fig.~\ref{fig:figd5}-Left, expanded in the cross-channel conformal partial waves. $C^{(0)}_{\m,J} = R^{1234}_{\m, J}$ is the ''OPE function" for this diagram \cite{Caron-Huot:2017vep}.

\subsection{Ladder diagram at 2-loop}
\label{sec:AA2}

We show in this section the details of the calculation of the 2-loop ladder diagram, which will then then easily generalise to $N$-loops. In particular, we will derive the following conformal partial wave expansion:
\begin{align}
\label{eq:fgk65s}
\mm{M}^{(2)}= \sum^\infty_{J_5 =0} \int_{-\infty}^{\infty}  \frac{d\m_5}{2\pi i} C^{(2)}_{\m_5,J_5} \Psi^{1234}_{\m_5, J_5}
\end{align}
and will explicitly derive the OPE function $C^{(2)}_{\m_5,J_5} $.

The 4-point ladder diagram at two-loops is given by (see Fig.~\ref{fig:figd5}-Right):
\begin{align}
\label{eq:fnndd}
&\mm{M}^{(2)}= \int d^{d+1} x_1 \cdots d^{d+1}x_6\  \nabla^L_1  \ K_{\D_1}(P_1,x_1) \nabla^L_2   \nabla_{x_2} K_{\D_2}(P_2,x_2) \times
\nn
&\nabla^L_3  K_{\D_3}(P_3,x_3) \nabla^L_4  K_{\D_4}(P_4,x_4)  \times
\nn
&G_{\D_5,L}(x_1,x_2) G_{\D_7,L}(x_3,x_4)\ G_{\D_{11},L}(x_5,x_6) \times
\nn
&\nabla^L_5  G_{\D_8} (x_1, x_5)\  \nabla^L_6  G_{\D_6} (x_2, x_6)\   G_{\D_9} (x_5, x_3)\   G_{\D_{10}} (x_6, x_4)~.
\end{align} 
Here we have defined $\nabla^L_i \equiv(K_i \cdot \nabla_{x_i})^L$. Now we will use the spectral representation for the four bulk-to-bulk propagators on the last line above:
\begin{align}
&\nabla^L_5  G_{\D_8} (x_1, x_5)  = \nabla^L_5  \int_{-\infty}^\infty d\n_8 \frac{1}{\n_8^2+(\D_8-\frac{d}{2})^2} \O_{\n_8}(x_1, x_5)
\nn
&\nabla^L_6  G_{\D_6} (x_2, x_6) =  \nabla^L_6 \int_{-\infty}^\infty d\n_6 \frac{1}{\n_6^2+(\D_6-\frac{d}{2})^2} \O_{\n_6}(x_2, x_6)
\end{align} 
and
\begin{align}
&G_{\D_9} (x_5, x_3)  =  \int_{-\infty}^\infty d\n_9 \frac{1}{\n_9^2+(\D_9-\frac{d}{2})^2} \O_{\n_9}(x_5, x_3)
\nn
&G_{\D_{10}} (x_6, x_4)  =  \int_{-\infty}^\infty d\n_{10} \frac{1}{\n_{10}^2+(\D_{10}-\frac{d}{2})^2} \O_{\n_{10}}(x_6, x_4)
\end{align} 
Now we use the split representation for the AdS harmonic functions:
\begin{align}
 \O_{\n_8}(x_1, x_5) = \frac{\n_8^2    }{\pi} \int d^dQ_8 K_{\frac{d}{2}-i\n_8 } (Q_8,x_1)   K_{\frac{d}{2}+i\n_8 } (Q_8,x_5)
 \nn
 \O_{\n_6}(x_2, x_6)  = \frac{\n_6^2  }{\pi} \int d^dQ_6 K_{\frac{d}{2}+i\n_6} (Q_6,x_2)  K_{\frac{d}{2}-i\n_6} (Q_6,x_6)
 \nn
\O_{\n_9}(x_5, x_3) = \frac{\n_9^2    }{\pi} \int d^dQ_9 K_{\frac{d}{2}-i\n_9 } (Q_9,x_5)K_{\frac{d}{2}+i\n_9 } (Q_9,x_3)
\nn
\O_{\n_{10}}(x_6, x_4) = \frac{\n_{10}^2    }{\pi} \int d^dQ_{10} K_{\frac{d}{2}-i\n_{10} } (Q_{10},x_6) K_{\frac{d}{2}+i\n_{10} } (Q_{10},x_4)
\end{align} 
Here the $Q_j$ are points on the boundary. Thus Eq.~\ref{eq:fnndd}  becomes:
\begin{align}
&\mm{M}^{(2)}=  \frac{1}{\pi^4}  \prod_{i=6,8,9,10} \int_{-\infty}^\infty d\n_i  \int d^dQ_i  \frac{ \n_i^2   }{ \n_i^2+(\D_i-\frac{d}{2})^2 } 
\nn
&\int d^{d+1}x_1 d^{d+1}x_2 \ \nabla^L_1   K_{\D_1}(P_1,x_1) \nabla^L_2 K_{\D_2}(P_2,x_2)    K_{\frac{d}{2}-i\n_8 } (Q_8,x_1)K_{\frac{d}{2}+i\n_6 } (Q_6,x_2)  G_{\D_5,L}(x_1,x_2)
\nn
&\int d^{d+1}x_3 d^{d+1}x_4 \ \nabla^L_3 K_{\D_3}(P_3,x_3) \nabla^L_4 K_{\D_4}(P_4,x_4) K_{\frac{d}{2}+i\n_9 } (Q_9,x_3)K_{\frac{d}{2}+ i\n_{10}} (Q_{10},x_4)   G_{\D_7,L}(x_3,x_4)
\nn
&\int d^{d+1}x_5 d^{d+1}x_6 \ \nabla^L_5 K_{\frac{d}{2}+i\n_8 }(Q_8,x_5)   \nabla^L_6 K_{\frac{d}{2}-i\n_6 }(Q_6,x_6) K_{\frac{d}{2}-i\n_{10} } (Q_{10},x_6) K_{\frac{d}{2}-i\n_{9} } (Q_{9},x_5)  G_{\D_{11},L}(x_5,x_6)
\end{align}
We see that we have a product of three tree-level exchange diagrams:
\begin{align}
\mm{M}^{(2)}= \frac{1}{\pi^4}  \prod_{i=6,8,9,10} \int_{-\infty}^\infty d\n_i  \int d^dQ_i  \frac{ \n_i^2   }{ \n_i^2+(\D_i-\frac{d}{2})^2 }\ \  \mm{M}^{6,2,1,\tilde{8}}_{5 , exch} \otimes  \mm{M}^{8, \tilde{6}, \tilde{10}, \tilde{9} }_{11 , exch} \otimes  \mm{M}^{3, 9, 10, 4 }_{7 , exch}
\end{align}
Now we use Eq.~\ref{eq:lkd3}, to expand the tree-level diagrams above in the crossed channel:
\begin{align}
&\mm{M}^{621\tilde{8}}_{\mm{O}_5 , exch} =\sum^\infty_{J_5 =0} \int \frac{d\m_5}{2\pi i}\ R^{\zz{12}6\tilde{8}}_{\mm{O}_5 ,O_{\m_5,J_5}}  \ \Psi^{126\tilde{8}}_{ \m_5,J_5}
\nn
&\mm{M}^{3,9,10,4}_{\mm{O}_7 , exch} =\sum^\infty_{J_7 =0} \int \frac{d\m_7}{2\pi i}\ R^{10,9, \zz{34}}_{\mm{O}_7 ,O_{\m_7,J_7}}  \ \Psi^{10,9,3,4}_{ \m_7,J_7}
\nn
&\mm{M}^{8,\tilde{6},\tilde{10},\tilde{9}}_{\mm{O}_{11}, exch} =\sum^\infty_{J_{11} =0} \int \frac{d\m_{11}}{2\pi i}\ R^{  \tilde{6},8 ,\tilde{9},\tilde{10}}_{\mm{O}_{11} ,O_{\m_{11},J_{11}}}  \ \Psi^{ \tilde{6},8 ,\tilde{9},\tilde{10}}_{ \m_{11},J_{11}}
\end{align}
Thus,
\begin{align}
\label{eq:fkdds}
&\mm{M}^{(2)} =\frac{1}{\pi^4} \int_{-\infty}^\infty  \bigg( \prod_{i=6,8,9,10}   d\n_i  \frac{ \n_i^2   }{ \n_i^2+(\D_i-\frac{d}{2})^2 } \bigg) \sum^\infty_{J_5 =0} \sum^\infty_{J_7 =0} \sum^\infty_{J_{11} =0} \int \frac{d\m_5}{2\pi i}  \int \frac{d\m_7}{2\pi i} \int \frac{d\m_{11}}{2\pi i}  
\nn
&R^{\zz{12}6\tilde{8}}_{\mm{O}_5 ,O_{\m_5,J_5}}   R^{ \tilde{6},8 ,\tilde{9},\tilde{10}}_{\mm{O}_{11} ,O_{\m_{11},J_{11}}} R^{10,9, \zz{34}}_{\mm{O}_7 ,O_{\m_7,J_7}}  
\times \int d^dQ_8 d^dQ_6  \int d^dQ_9 d^dQ_{10} \Psi^{126\tilde{8}}_{ \m_5,J_5}    \Psi^{10,9,3,4}_{ \m_7,J_7}  \Psi^{ \tilde{6},8 ,\tilde{9},\tilde{10}}_{ \m_{11},J_{11}}
\end{align}
The CPWs above have a shadow representation:
\begin{align}
\label{eq:pol266}
&\Psi^{126\tilde{8}}_{ \m_5,J_5}    =  \int d^dP_0 \langle O(P_1)O(P_2) O_{\m_5, J_5}(P_0 ) \rangle  \langle  \tilde{O}_{\m_5, J_5}(P_0)\tilde{O}(Q_8)O(Q_6) \rangle
\nn
&\Psi^{10,9,3,4}_{ \m_7,J_7} =  \!\!\int d^dP_0' \langle O(Q_{10}) O(Q_9) O_{\m_7, J_7}(P_0 ') \rangle  \langle  \tilde{O}_{\m_7, J_7}(P_0 ')O(P_3)O(P_4) \rangle
\nn
&\Psi^{\tilde{6},8 ,\tilde{9},\tilde{10}}_{ \m_{11},J_{11}} =  \!\!\int d^dP_0'' \langle \tilde{O}(Q_{6}) O(Q_8) O_{\m_{11}, J_{11}}(P_0 '') \rangle  \langle  \tilde{O}_{\m_{11}, J_{11}}(P_0 '')\tilde{O}(Q_9) \tilde{O}(Q_{10}) \rangle
\end{align} 
Now we plug Eq.~\ref{eq:pol266} in Eq.~\ref{eq:fkdds}, and we can do the $Q_6$, $Q_8$ integrals\footnote{The factor $B_{\m_5,J_5}$ is a known function, defined e.g in appendix A of \cite{Meltzer:2019nbs}}:
\begin{align}
\label{eq:pol4}
&\int d^dQ_8 d^dQ_6 \langle  \tilde{O}_{\m_5, J_5}(P_0)\tilde{O}(Q_8)O(Q_6) \rangle \langle  O(Q_8)\tilde{O}(Q_6) O_{\m_{11}, J_{11}}(P_0 '') \rangle
\nn
&= B_{\m_5,J_5} \d_{P_0,P_0''} \d_{\m_5,\m_{11}} \d_{J_5,J_{11}}~.
\end{align} 
and the $Q_9$, $Q_{10}$ integrals as follows:
\begin{align}
\label{eq:pol4563}
&\int d^dQ_9 d^dQ_{10} \langle O(Q_{10}) O(Q_9) O_{\m_7, J_7}(P_0 ') \rangle  \langle  \tilde{O}_{\m_{11}, J_{11}}(P_0 '')\tilde{O}(Q_9) \tilde{O}(Q_{10}) \rangle
\nn
&= B_{\m_7,J_7} \d_{P_0',P_0''} \d_{\m_7,\m_{11}} \d_{J_7,J_{11}}~.
\end{align} 
Thus:
\begin{align}
 \int d^dQ_8 d^dQ_6  \int d^dQ_9 d^dQ_{10} \Psi^{126\tilde{8}}_{ \m_5,J_5}    \Psi^{10,9,3,4}_{ \m_7,J_7}  \Psi^{ \tilde{6},8 ,\tilde{9},\tilde{10}}_{ \m_{11},J_{11}}
=B_{\m_5,J_5} B_{\m_7,J_7}    \Psi^{1234}_{\m_5, J_5} \d_{\m_5,\m_7} \d_{J_5,J_7} d_{\m_7,\m_{11}} \d_{J_7,J_{11}}
\end{align} 
and we get:
\begin{align}
 \frac{1}{\pi^4}    \sum^\infty_{J_5 =0} \int \frac{d\m_5}{2\pi i}  \bigg[ (B_{\m_5,J_5})^2   \int \Big( \prod_{i=6,8,9,10}   \frac{ d\n_i \n_i^2   }{ \n_i^2+(\D_i-\frac{d}{2})^2 }  \Big)  R^{\zz{12}6\tilde{8}}_{\mm{O}_5 ,O_{\m_5,J_5}}   R^{ \tilde{6},8 ,\tilde{9},\tilde{10}}_{\mm{O}_{11} ,O_{\m_{11},J_{11}}} R^{10,9, \zz{34}}_{\mm{O}_7 ,O_{\m_7,J_7}}  \bigg] \Psi^{1234}_{\m_5, J_5}
\end{align}
For clarity, we highlighted in red the external operators in $R$. The factor in the square brackets above gives the OPE function:
\begin{align}
C^{(2)}_{\m_5,J_5}  = \frac{(B_{\m_5,J_5})^2}{\pi^4}  \int_{-\infty}^\infty \Big( \prod_{i=6,8,9,10}   \frac{ d\n_i \n_i^2   }{ \n_i^2+(\D_i-\frac{d}{2})^2 }  \Big)  R^{\zz{12}6\tilde{8}}_{\mm{O}_5 ,O_{\m_5,J_5}}      R^{ \tilde{6},8 ,\tilde{9},\tilde{10}}_{\mm{O}_{11} ,O_{\m_{11},J_{11}}} R^{10,9, \zz{34}}_{\mm{O}_7 ,O_{\m_7,J_7}}  
\end{align}
And we derived Eq.~\ref{eq:fgk65s}:
\begin{align}
\mm{M}^{(2)}= \sum^\infty_{J_5 =0} \int_{-\infty}^{\infty} \frac{d\m_5}{2\pi i} C^{(2)}_{\m_5,J_5} \Psi^{1234}_{\m_5, J_5}
\end{align}




\section{Diagrams in terms of sums of Jacobi/Legendre functions}
\label{sec:B1}

The conformal block in $d=2$ is:
\begin{align} 
\label{eq:cnd2}
\mathcal{K}^{\D_i}_{\D,l} (z,\bar z) = \frac{1}{1+\d_{l,0}} \Big[k_{\D-l}(z)k_{\D+l}(\bar z) +k_{\D+l}( z)k_{\D-l}(\bar z) \Big]
\end{align} 
and in $d=4$:
\begin{align} 
\label{eq:cnd4}
\mathcal{K}^{\D_i}_{\D,l} (z,\bar z) = \frac{z\bar z}{\bar z-z} \Big[k_{\D-l-2}(z)k_{\D+l}(\bar z) -k_{\D+l}( z)k_{\D-l-2}(\bar z) \Big]
\end{align} 
We can write the conformal blocks in terms of Jacobi functions of the second kind, using the following relation:
\begin{align} 
k_\b(z) \equiv z^{\frac{\b}{2}} {}_2F_1 (\frac{\b}{2}+a,\frac{\b}{2}+b,\b,z) 
=  \frac{(-1)^bz^a}{(z-1)^{a+b}} \frac{2\G_\b}{\G_{\frac{\b}{2}-b} \G_{\frac{\b}{2}+b} } Q^{(b-a,-b-a)}_{\frac{\b}{2}+a-1} (\hat z)
\end{align} 
Then the 2d conformal block Eq.~\ref{eq:cnd2} becomes:
\begin{align} 
&\mathcal{K}^{\D_i}_{\frac{d}{2}+i\nu,l} (z,\bar z) =  \frac{4}{1+\d_{l,0}} (z\bar z)^a ((z-1)(\bar z-1))^{-a-b}  
\nn
&\times \frac{\G_{2s+l+1} \G_{2s-l+1} }{\G_{s+\frac{l}{2}-b+\frac{1}{2}}\G_{s+\frac{l}{2}+b+\frac{1}{2}}\G_{s-\frac{l}{2}+b+\frac{1}{2}}  \G_{s-\frac{l}{2}-b+\frac{1}{2}}}  Q^{(b-a,-b-a)}_{s+\frac{l}{2}+a-\frac{1}{2}} (\hat z) Q^{(b-a,-b-a)}_{s-\frac{l}{2}+a-\frac{1}{2}} (\hat{\bar z}) +(z \leftrightarrow \bar z)
\nn
\end{align} 
where $s\equiv \frac{i\n}{2}$. The 4d conformal block Eq.~\ref{eq:cnd4} becomes:
\begin{align} 
&\mathcal{K}^{\D_i}_{\frac{d}{2}+i\nu,l} (z,\bar z) =  \frac{4z\bar z}{\bar z-z}(z\bar z)^a ((z-1)(\bar z-1))^{-a-b}  
\nn
&\frac{\G_{2s+l+2}}{\G_{s+\frac{l}{2}+1-b} \G_{s+\frac{l}{2}+1+b} }   \frac{\G_{2s-l}}{\G_{s-\frac{l}{2}-b} \G_{s-\frac{l}{2}+b} }  Q^{(b-a,-b-a)}_{s+\frac{l}{2}+a} (\hat z)  Q^{(b-a,-b-a)}_{s-\frac{l}{2}+a-1} (\hat{\bar z}) +(z \leftrightarrow \bar z)
\end{align} 
We can now plug this in Eq.~\ref{eq:98d3} to compute 4-point ``Blob diagrams" in terms integrals/sums of Jacobi functions.

\subsection{$b=0$, general $a$}

For simplicity, let us focus on the case of $b=0$ and general $a$, where the blocks can be written in terms of associated Legendre functions of the second kind, using the following relation:
 \begin{align} 
k_\b(z) =  \frac{2^{\b}}{\sqrt{\pi}}\frac{e^{i\pi a}}{\G_{\frac{\b}{2}-a}}\G_{\frac{\b}{2}+\frac{1}{2}} (1-z)^{\frac{-a}{2}} Q^{-a}_{\frac{\b}{2}-1} (\hat{z}) =2e^{i\pi a} \frac{\G_{\b}}{\G_{\frac{\b}{2}}\G_{\frac{\b}{2}-a}} (1-z)^{\frac{-a}{2}} Q^{-a}_{\frac{\b}{2}-1} (\hat{z})
\end{align} 
where $\hat z = \frac{2}{z}-1$. The $d=2$ conformal block Eq.~\ref{eq:cnd2} becomes:
\begin{align} 
\label{eq:dfkj65}
&\mathcal{K}^{\D_i}_{\frac{d}{2}+i\nu,l} (z,\bar z)=\frac{1}{1+\d_{l,0}}  4e^{2 i\pi a} ((1-z)(1-\bar z))^{\frac{-a}{2}}  \times
\nn
&\Big[ \frac{\G_{1+2s+l}}{\G_{\frac{1+2s+l}{2}}\G_{\frac{1+2s+l}{2}-a}}    \frac{\G_{1+2s-l}}{\G_{\frac{1+2s-l}{2}}\G_{\frac{1+2s-l}{2}-a}} Q^{-a}_{\frac{2s+l}{2}-\frac{1}{2}} (\hat{z}) Q^{-a}_{\frac{2s-l}{2}-\frac{1}{2}} (\hat{\bar{z}} )  \Big] +(z \leftrightarrow \bar z)
\end{align} 
where $s\equiv \frac{i\n}{2}$. Then the scalar $l=0$ block is:
\begin{align} 
\mathcal{K}^{\D_i}_{\frac{d}{2}+i\nu,l=0} (z,\bar z)= \frac{e^{2 \pi i a}}{\pi}((1-z)(1-\bar z))^{\frac{-a}{2}} 4^{1+2s}\frac{\G_{s+1}^2}{\G_{s-a+\frac{1}{2}}^2} Q^{-a}_{s-\frac{1}{2}} (\hat{z})Q^{-a}_{s-\frac{1}{2}} (\hat{\bar z})
\end{align} 
Using Eq.~\ref{eq:sdskkf} with $b=0$ gives:
\begin{align} 
&\Upsilon_\nu^{\D_i} \times \mathcal{K}^{\D_i}_{\frac{d}{2}+i\nu} (z,\bar z) = 4 e^{2 \pi i a}((1-z)(1-\bar z))^{\frac{-a}{2}}  \times
\nn
&\Big( \frac{ \G_{s-\frac{1}{2}+a_p} \G_{s-\frac{1}{2}+b_p}}{ \G_{s+\frac{3}{2}-a_p} \G_{s+\frac{3}{2}-b_p} } \Big) \frac{\G_{s+\frac{1}{2}+a}}{\G_{s+\frac{1}{2}-a}}  (2s) Q^{-a}_{s-\frac{1}{2}} (\hat{z})Q^{-a}_{s-\frac{1}{2}} (\hat{\bar z})
\end{align} 
Using the identity $Q^{-a}_{s-\frac{1}{2}} (\hat{\bar z}) = e^{-2 \pi i a} \frac{\G_{s+\frac{1}{2}-a}}{\G_{s+\frac{1}{2}+a}} Q^{a}_{s-\frac{1}{2}} (\hat{\bar z})$, we get:
\begin{align} 
&\Upsilon_\nu^{\D_i} \times \mathcal{K}^{\D_i}_{\frac{d}{2}+i\nu} (z,\bar z) = 4  ((1-z)(1-\bar z))^{\frac{-a}{2}}  \times
\nn
&\Big( \frac{ \G_{s-\frac{1}{2}+a_p} \G_{s-\frac{1}{2}+b_p}}{ \G_{s+\frac{3}{2}-a_p} \G_{s+\frac{3}{2}-b_p} } \Big)  (2s)  Q^{-a}_{s-\frac{1}{2}} (\hat{z})Q^{a}_{s-\frac{1}{2}} (\hat{\bar z})
\end{align} 
Thus, from Eq.~\ref{eq:98d3} we have the expression for the $l=0$ blob diagram:
\begin{align}
\label{eq:cnss8d} 
g_{l=0}(z,\bar z) = 
4\pi^2 \int d\n \tilde{F}_\n  \frac{ \Big( \frac{ \G_{\frac{i\n}{2}-\frac{1}{2}+a_p} \G_{\frac{i\n}{2}-\frac{1}{2}+b_p}}{ \G_{\frac{i\n}{2}+\frac{3}{2}-a_p} \G_{\frac{i\n}{2}+\frac{3}{2}-b_p} } \Big) }{\sin \pi(a_p-\frac{1}{2}-\frac{i\n}{2}) \sin \pi(b_p-\frac{1}{2}-\frac{i\n}{2})   } (i\n) Q^{-a}_{\frac{i \n-1}{2}} (\hat z) Q^a_{\frac{i \n-1}{2}} (\hat{\bar{z}})
\nn 
\end{align}

\subsection{$a=b=0$}

Let us now look at the case when both $a$ and $b$ are zero. From Eq.~\ref{eq:cnss8d}, we have for $l=0$:
\begin{align} 
g_{l=0}(z,\bar z) = 
4\pi^2 \int d\n \tilde{F}_\n  \frac{ \frac{\G_{\D_1-\frac{1}{2} -\frac{i\n}{2}} \G_{\D_4-\frac{1}{2} -\frac{i\n}{2}}} { \G_{-\D_1+\frac{3}{2} +\frac{i\n}{2}}\G_{-\D_4+\frac{3}{2} +\frac{i\n}{2}}} }{\sin \pi(\D_1-\frac{1}{2}-\frac{i\n}{2}) \sin \pi(\D_4-\frac{1}{2}-\frac{i\n}{2})  } (i\n) Q_{\frac{i \n-1}{2}} (\hat z) Q_{\frac{i \n-1}{2}} (\hat{\bar{z}})
\nn 
\end{align} 
From Eq.~\ref{eq:dfkj65} and for $l=1$:
\begin{align} 
g_{l=1}(z,\bar z) = 
\pi^2 \int d\n \tilde{F}_\n  \frac{ \frac{\G_{\D_1 -\frac{i\n}{2}} \G_{\D_4 -\frac{i\n}{2}}} { \G_{-\D_1+1 +\frac{i\n}{2}}\G_{-\D_4+1 +\frac{i\n}{2}}} }{\sin \pi(\D_1 -\frac{i\n}{2}) \sin \pi(\D_4 -\frac{i\n}{2})  } (i\n) Q_{\frac{i \n}{2}} (\hat z) Q_{\frac{i \n}{2}-1} (\hat{\bar{z}}) +(z \leftrightarrow \bar z)
\nn 
\end{align} 
Using the identity:
\begin{align} 
\Big[ \frac{1-\hat{z}^2}{\frac{i \n}{2}}\pa_{\hat z} +\hat z \Big] Q_{\frac{i \n}{2}} (\hat z) =Q_{\frac{i \n}{2}-1} (\hat z)
\end{align} 
we get:
\begin{align} 
g_{l=1}(z,\bar z) = 
\pi^2 \int d\n \tilde{F}_\n  \frac{ \frac{\G_{\D_1 -\frac{i\n}{2}} \G_{\D_4 -\frac{i\n}{2}}} { \G_{-\D_1+1 +\frac{i\n}{2}}\G_{-\D_4+1 +\frac{i\n}{2}}} }{\sin \pi(\D_1 -\frac{i\n}{2}) \sin \pi(\D_4 -\frac{i\n}{2})  } (i\n)  \Big[ \frac{1-\hat{z}^2}{\frac{i \n}{2}}\pa_{\hat z} +\hat z \Big]  Q_{\frac{i \n}{2}} (\hat z) Q_{\frac{i \n}{2}} (\hat{\bar{z}}) +(z \leftrightarrow \bar z)
\nn 
\end{align} 
Now for $\D_1=\D_4=1$
\begin{align} 
g_{l=1}(z,\bar z) = 
-\frac{i}{4}\pi^2 \int d\n \tilde{F}_\n  \frac{\n^3}{\sin \pi(1 -\frac{i\n}{2})^2  }  \Big[ \frac{1-\hat{z}^2}{\frac{i \n}{2}}\pa_{\hat z} +\hat z \Big]  Q_{\frac{i \n}{2}} (\hat z) Q_{\frac{i \n}{2}} (\hat{\bar{z}}) +(z \leftrightarrow \bar z)
\nn 
\end{align}
Now we can close the contour and use the residue theorem find expressions for the blob diagrams in position space. The sums that one gets are nearly canonical sums of Legendre functions, and one can hope to get analytical expressions. We leave this for future work.

\section{More relations for 4-point blob diagrams}
\label{sec:se5}

\subsection{4-point functions with $a=0$ and $b=\frac{1}{2}$}
\label{sec:c1}

\begin{figure}[t]
\centering
\includegraphics[clip,height=3.3cm]{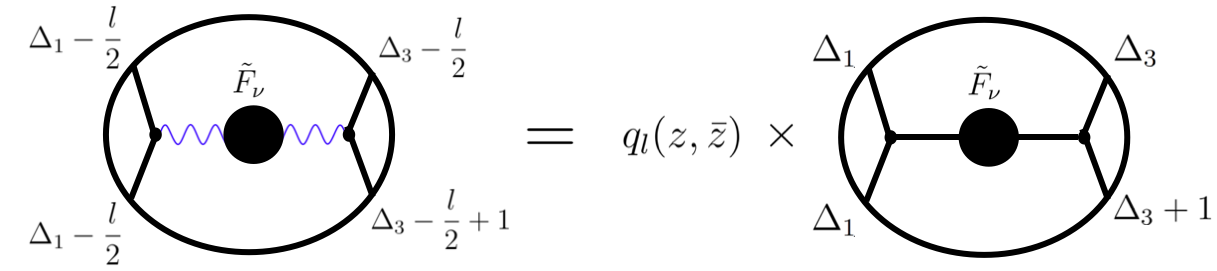}
\caption{Showing the relation of Eq.~\ref{eq:vcnmsfd6} for the 4-point function, for $a=0$ and $b=\frac{1}{2}$ and $d=2$. The general ``blob" $\tilde{F}_\n$ is the same on both sides. On the left we have an exchange of spin-$l$, and on the right we have exchange of spin-0.}\label{fig:lfkr4} 
\end{figure}

In this section we derive a relation between the 4-point ``blob diagrams" with different spin-$l$ in $d=2$ dimensions. We start with expression Eq.~\ref{eq:98d3}. The conformal block in $d=2$ is:
\begin{align} 
\mm{K}_{\D,l} (z,\bar z)  = \frac{1}{1+\d_{l,0}} \Big(k_{\D-l}^{(a,b)}(z) k_{\D+l}^{(a,b)}(\bar z) + k_{\D+l}^{(a,b)}(z) k_{\D-l}^{(a,b)}(\bar z) \Big)
\end{align} 
where $k_\b^{(a,b)}(z) \equiv z^{\frac{\b}{2}} {}_2 F_1 (\frac{\b}{2}+a,\frac{\b}{2}+b,\b,z)$. For $a=0$ and $b=\frac{1}{2}$, the $d=2$ conformal block simplifies, the spin-$l$ block is proportional to the spin-0 block:
\begin{align} 
\mm{K}_{1+i\n,l} (z,\bar z) =  \Big( \frac{\sqrt z(1+\sqrt{1-\bar z})}{4\sqrt{\bar z} (1+\sqrt{1- z})} \Big)^{l} \mm{K}_{1+i\n,l=0} (z,\bar z) 
\end{align} 
Where the the scalar $l=0$ block for $a=0$ and $b=\frac{1}{2}$ is:
\begin{align} 
\label{eq:cnns79}
\mathcal{K}^{\D_i}_{\frac{d}{2}+i\n, l=0 }(z,\bar z)   = \sqrt{\frac{u}{v}}   \big(4Z \big)^{i\n}
\end{align} 
where $Z\equiv \frac{\sqrt{z \bar z}}{(1+\sqrt{1-z})(1+\sqrt{1-\bar z})}$. From Eq.~\ref{eq:sdskkf}, the function $\Upsilon_\nu^{\D_i}$ for any $d$ and for $a=0$ and $b=\frac{1}{2}$, obeys:
\begin{align} 
 \Upsilon_{\nu,l} |_{a_{p}\to a_p-\frac{l}{2},b_{p}\to b_p-\frac{l}{2}} = 4^{-l}\Upsilon_{\nu,l=0} 
\end{align} 
Combining the previous two equations, we have
\begin{align} 
 \Upsilon_{\nu,l}  \mm{K}_{1+i\n,l} (z,\bar z)|_{a_{p}\to a_p-\frac{l}{2},b_{p}\to b_p-\frac{l}{2}}  = q_l(z,\bar z) { \Upsilon_{\nu,l=0}^{\D_i}\mm{K}_{1+i\n,l=0 } (z,\bar z)} 
\end{align} 
where $ q_l(z, \bar z) \equiv \Big( \frac{\sqrt z(1+\sqrt{1-\bar z})}{16\sqrt{\bar z} (1+\sqrt{1- z})} \Big)^{l}$. Thus we get from Eq.~\ref{eq:98d3} for the 4-point function of a ''blob diagram":
\begin{align}
\label{eq:vcnmsfd6}
\boxed{
g_l(z,\bar z)|_{a_{p}\to a_p-\frac{l}{2},b_{p}\to b_p-\frac{l}{2}} = q_l(z, \bar z) g_{l=0}(z,\bar z) }
\end{align}
where it is assumed that both sides of this equation have the same $F_\n$ spectral function (Eq.~\ref{eq:98d3}), but otherwise this function is general. We thus see that in $d=2$ and $a=0$ and $b=\frac{1}{2}$ the 4-point ``blob diagrams" with any $l$ can be directly computed from an $l=0$ diagram. This relation is illustrated in Fig.~\ref{fig:lfkr4}.

Using identity 0 Eq.~\ref{eq:nbvl3}, we can raise and lower $a$ and $b$, and hence relate 2d diagrams with $a$=integer and $b=$half-integer. Additionally, we expect that in $d=4$ dimensions there should be a similar relation to Eq.~\ref{eq:vcnmsfd6}, but we leave this to future work.

\subsection{A relation across dimensions $d$}
\label{sec:c2}

Let us recall the expression for a ''Blob diagram" Eq.~\ref{eq:98d3}:
\begin{align} 
\label{eq:3487f}
 g^{(\D_i)}_{l,d}(z,\bar z)  = \int d\nu 
\tilde{F}_{\n,l} \times T_{\nu,l,d}^{(\D_i)}  \times \mathcal{K}^{\D_i}_{\frac{d}{2}+i\nu,l} (z,\bar z) 
\end{align}
where we defined
\begin{align} 
\label{eq:fdmnf4}
T_{\nu,l,d}^{(\D_i)} \equiv \frac{ \Upsilon_{\nu,l}^{\D_i}  }{  \sin \pi (\frac{\Delta_1+\D_2+l}{2} -\frac{d}{4}-\frac{i \nu}{2}) \sin \pi (\frac{\Delta_3+\D_4+l}{2} -\frac{d}{4}-\frac{i \nu}{2})}  
\end{align} 
$\D_i$ are the external scaling dimensions. Recall that the conformal blocks in $d=2$ and $d=4$ are a symmetrized sum in $z$, $\bar z$:
\begin{align} 
\mm{K}^{\D_i}_{\D,l,d=2} (z,\bar z)  = \frac{1}{1+\d_{l,0}} \Big(k_{\D-l}^{(a,b)}(z) k_{\D+l}^{(a,b)}(\bar z) + k_{\D+l}^{(a,b)}(z) k_{\D-l}^{(a,b)}(\bar z) \Big)
\nn
\mm{K}^{\D_i}_{\D,l,d=4} (z,\bar z)  = \frac{z\bar z}{z-\bar z} \Big(k_{\D-l-2}^{(a,b)}(z) k_{\D+l}^{(a,b)}(\bar z) - k_{\D+l}^{(a,b)}(z) k_{\D-l-2}^{(a,b)}(\bar z) \Big)
\end{align} 
Thus we can write Eq.~\ref{eq:3487f} as:
\begin{align}
 g^{(\D_i)}_{l,d}(z,\bar z)  =  \int_{-\infty}^{\infty} d\n\  \widetilde{g}^{(\D_i)}_{\n, l,d}(z,\bar z)  + \int_{-\infty}^{\infty} d\n\ \widetilde{g}^{(\D_i)}_{\n, l,d}(\bar z, z)
\end{align}
where
\begin{align}
\label{eq:vndf4df1}
\widetilde{g}^{(\D_i)}_{\n,l,d=2}(z,\bar z) \equiv  \frac{1}{1+\d_{l,0}}  
\tilde{F}_{\n,l} \times T_{\nu,l,d=2}^{(\D_i)}   \times k_{i\n+1-l}^{(a,b)}(z) k_{i\n+1+l}^{(a,b)}(\bar z) 
\end{align}
and
\begin{align}
\label{eq:vndf4df2}
 \widetilde{g}^{(\D_i)}_{\n,l,d=4}(z,\bar z) \equiv \frac{z\bar z}{z-\bar z}  
\tilde{F}_{\n,l} \times T_{\nu,l,d=4}^{(\D_i)}  \times k_{i\n-l}^{(a,b)}(z) k_{i\n+2+l}^{(a,b)}(\bar z) 
\end{align}
Now we will find a relation between $\widetilde{g}^{(\D_i)}_{\n,l,d=2}(z,\bar z) $ and $\widetilde{g}^{(\D_i)}_{\n,l,d=4}(z,\bar z) $, Eqs.~\ref{eq:vndf4df1} and \ref{eq:vndf4df2}. Consider the following transformation:
\begin{align} 
\label{eq:symmery2}
(l,d,\D_i) \longrightarrow (l-1,d+2,\D_i+1) 
\end{align} 
This is a transformation that simultaneously changes the spin, the space-time dimension, and the scaling dimensions of a blob diagram. Under this transformation we have (Eq.~\ref{eq:sdskkf} and \ref{eq:fdmnf4}):
\begin{align} 
T_{\nu,l-1,d=4}^{(\D_i+1)} =  (1+\d_{l,0}) \frac{z\bar z}{z-\bar z}   (i \n )   T_{\nu,l,d=2}^{(\D_i)} 
\end{align} 
and thus we get the relation:
\begin{align} 
\boxed{
 \widetilde{g}_{\nu,l-1,d=4}^{(\D_i+1)}(z,\bar z) =  (1+\d_{l,0}) \frac{z\bar z}{z-\bar z} \times  (i \n )   \widetilde{g}_{\nu,l,d=2}^{(\D_i)} (z,\bar z) }
\end{align} 
This relation connects ``blob diagrams" in $d=2$ and $d=4$. In $d=6$ the conformal block is likewise a sum of product of functions $k_{\D}^{(a,b)}(z)$ \cite{Dolan:2011dv}. We expect to have a relation connecting blob diagrams in $d=6$ and $d=2, 4$. We will leave this for future work.

\bibliographystyle{utphys}
\bibliography{Arxiv}

\end{document}